\documentclass[journal,onecolumn,12pt,draftclsnofoot]{IEEEtran}
\ifCLASSINFOpdf
\else
\fi
\usepackage{epsfig,graphics,graphicx,subfig,amssymb,amstext,amsmath,algorithm,algorithmic,wrapfig,multirow}
\usepackage{array}
\usepackage{cite}
\usepackage{url}
\usepackage{times}
\usepackage{lipsum}
\usepackage{float}
\usepackage{placeins}
\usepackage{textcomp}
\usepackage[font=small]{caption}
\usepackage{hyperref}
\usepackage{flushend}
\usepackage{xcolor}
\usepackage{tabularx}
\usepackage{bm}
\usepackage{enumerate}
\usepackage{tikz}
\usepackage{pgfplots}
\usepackage{enumerate}
\usepackage{epstopdf}
\usetikzlibrary{shapes,arrows}
\usepackage[utf8]{inputenc}
\usepackage{mathtools}
\usepackage{color}

\def\beq{\begin{equation}}
\def \eeq{\end{equation}}
\def\beqa{\begin{eqnarray}}
\def\eeqa{\end{eqnarray}}
\def\beqan{\begin{eqnarray*}}
\def\eeqan{\end{eqnarray*}}

\def\arr{\rightarrow}
\def\Exp{\mathbb{E}}

\def\tm1{t\! - \! 1}
\def\tp1{t\! + \! 1}

\def\nbf{\mathbf{n}}

\def\ubf{\mathbf{u}}

\def\vbf{\mathbf{v}}

\def\wbf{\mathbf{w}}

\def\ybf{\mathbf{y}}
\def\zbf{\mathbf{z}}

\def\Hbf{\mathbf{H}}

\def\Ibf{\mathbf{I}}

\def\Qbf{\mathbf{Q}}

\newif\ifconf
\conffalse

\newif\ifthreebitq
\threebitqtrue
\newif\ifonecol
\onecolfalse

\renewcommand{\footnoterule}{
  \kern -3pt
  \hrule width \columnwidth height 0.5pt
  \kern 3pt
}

\begin{document}

\title{A Case for Digital Beamforming at mmWave}

%\title{Quantifying the Low in Low Resolution Digital Beamforming for mmWave Cellular}


\author{Sourjya Dutta,~\IEEEmembership{Student Member,~IEEE}, C. Nicolas Barati,~\IEEEmembership{Student Member,~IEEE} David Ramirez,~\IEEEmembership{Member,~IEEE}, Aditya Dhananjay, James F. Buckwalter,~\IEEEmembership{Senior Member,~IEEE}  and Sundeep Rangan,~\IEEEmembership{Fellow,~IEEE}% <-this % stops a space
\thanks{The authors were supported in part by NSF Grants 1116589, 1302336, and 1547332, NIST award 70NANB17H166, SRC, and the affiliate members of NYU WIRELESS. S. Dutta, C.N. Barati, A. Dhananjay and S. Rangan are with NYU WIRELESS, Tandon School of Engineering, New York University, Brooklyn, NY 11201, USA (email: sdutta@nyu.edu, nicolas.barati@nyu.edu, aditya@courant.nyu.edu, srangan@nyu.edu). D. Ramirez is with the Department of Electrical Engineering, Princeton University, Princeton, NJ 08544 USA (e-mail: dard@princeton.edu). J.F. Buckwalter is with the Department of Electrical and Computer Engineering, University of California at Santa Barbara, Santa Barbara, CA 93106 USA (e-mail: buckwalter@ece.ucsb.edu). }
}
\maketitle

\begin{abstract}
Due to the heavy reliance of millimeter-wave (mmWave) wireless systems on directional links, beamforming (BF) with high-dimensional arrays is essential for cellular systems in these frequencies. How to perform the array processing in a power efficient manner is a fundamental challenge.
Analog and hybrid BF require fewer analog-to-digital and digital-to-analog converters (ADCs and DACs), but can only communicate in a small number of directions at a time, limiting directional search, spatial multiplexing and control signaling. Digital BF enables flexible spatial processing, but must be operated  at a low quantization resolution to stay within reasonable power levels. This decrease in quantizer resolution introduces noise in the received signal and degrades the quality of the transmitted signal. To assess the effect of low-resolution quantization on cellular system, we present a simple additive white Gaussian noise (AWGN) model for quantization noise. Simulations with this model reveal that at moderate resolutions (3-4 bits per ADC), there is negligible loss in downlink cellular capacity from quantization. In essence, the low-resolution ADCs limit the high SNR, where cellular systems typically do not operate. For the transmitter, it is shown that DACs with $4$ or more bits of resolution do not violate the adjacent carrier leakage limit set by 3\textsuperscript{rd} Generation Partnership Project (3GPP) New Radio (NR) standards for cellular operations. Further, this work studies the effect of low resolution quantization on the error vector magnitude (EVM) of the transmitted signal.In fact, our findings suggests that low-resolution fully digital BF architectures can be a power efficient alternative to analog or hybrid beamforming for both transmitters and receivers at millimeter wave.
%The findings suggest that low-resolution fully digital BF architectures can be power efficient, offer greatly enhanced control plane functionality and comparable data plane performance to analog BF.

%\textcolor{blue}{Millimeter wave (mmWave) communication networks could achieve the goals of 5th Generation Networks, but a beamforming architecture that is efficient in terms of, both, power and utilization must be developed to fully exploit the mmWave potential. Analog and hybrid beamforming architectures are efficient in terms of power consumption, but can only transmit in a single direction per stream. Alternatively, digital beamforming can transmit in a plethora of directions simultaneously, but can be power inefficient. To become power efficient, a digital beamforming architecture can utilize low bit resolution converters which introduce potentially harmful signal distortion. We characterize the trade-off between resolution and SINR for a mmWave cellular network using various multiple access techniques. Our results show that 4-bit and 2-bit resolution at the base station and user equipment, respectively, are sufficient to achieve ?. Furthermore, our analysis shows that when doing simultaneous orthogonal transmissions with digital beamforming, the noise due to signal distortion has a nearly negligible impact on the SINR.}

%In this work we assess the multiple access techniques available at the base station (BS) transmitted with fully digital beamforming using low resolution quantizers at millimeter frequencies.
\end{abstract}

\begin{IEEEkeywords}
Millimeter wave, 5G cellular, Low resolution quantizers, Digital beamforming.
\end{IEEEkeywords}

\IEEEpeerreviewmaketitle

\section{Introduction}
The need for more bandwidth, driven by ever higher demand, has brought millimeter wave (mmWave) communication into the spotlight as an enabling technology for the 5\textsuperscript{th} generation (5G) wireless communication systems. By offering large blocks of contiguous spectrum, mmWave presents a unique opportunity to overcome the bandwidth crunch problem in lower frequency bands \cite{RanRapE:14}.
At mmWave frequencies, high isotropic path loss necessitates the reliance on antenna arrays with large number of elements. These arrays overcome the path loss by high directional gains through beamforming (BF). Thus, a transmitter--receiver (Tx--Rx) pair uses large number of antennas to focus energy in a particular direction to meet a target link budget. A key challenge for large antenna arrays, and the motivation of our work, is to find an architecture capable of high-dimensional array processing in a power-efficient manner at mmWave frequencies.

\begin{figure}[ht!]
\centering
\begin{tikzpicture}[scale=0.5]
%\draw[thick] (0,0) rectangle (2,6) node [align=center,rotate=90] at (1,3.0) {$\mathbf{w_{\tiny TX}}$};
%\draw[thick] (18,0) rectangle (20,6) node [align=center,rotate=270] at (19,3.0) {$\mathbf{w_{\tiny RX}}$};

\foreach \y in {1,5} {

\draw[thick] (4,\y) -- (5,\y); \draw[thick] (16,\y) -- (15,\y);
\draw[thick] (5.5,\y) circle (0.5) node [align=center] at (5.5,\y) {\Huge $\nearrow$}; 
\draw[thick] (14.5,\y) circle (0.5) node [align=center] at (14.5,\y) {\Huge $\nearrow$}; 
\draw[thick] (6,\y) -- (7,\y); \draw[thick] (14,\y) -- (13,\y);
\draw[thick] (7,\y-0.5) rectangle (8,\y+0.5) node [align=center] at (7.5, \y) {\tiny PA};
\draw[thick] (13,\y-0.5) rectangle (12,\y+0.5) node [align=center] at (12.5, \y) {\tiny LNA};
\draw[thick] (8,\y) -- (8.5,\y); \draw[thick] (12,\y) -- (11.5,\y);

\draw[thick] (8.5,\y) -- (8.6,\y+0.1) -- (8.6,\y-0.1) --cycle;
\draw[thick] (11.5,\y) -- (11.4,\y+0.1) -- (11.4,\y-0.1) --cycle;
}

\draw[thick] (0,3) -- (1,3); 
\draw[thick] (-1,3-0.5) rectangle (0,3+0.5) node [align=center] at (-0.5, 3) {\tiny D/A};
\draw[thick] (1.5,3) circle (0.5) node [align=center] at (1.5,3) {$\times$}; 

\draw[thick] (20,3) -- (19,3); 
\draw[thick] (21,3-0.5) rectangle (20,3+0.5) node [align=center] at (20.5, 3) {\tiny A/D};
\draw[thick] (18.5,3) circle (0.5) node [align=center] at (18.5,3) {$\times$}; 

\draw[thick] (2,3) -- (3,3); \draw[thick] (17,3) -- (18,3);
\draw[thick] (4,5) -- (3.75, 3.5); \draw[thick] (4,1) -- (3.75, 2.5); 
\draw[thick] (16,5) -- (16.25, 3.5); \draw[thick] (16,1) -- (16.25, 2.5); 
\draw[thick] (3,2.5) rectangle (4,3.5) node [align=center] at (3.5,3) { \tiny{1:N}}; 
\draw[thick] (16.5,3) circle (0.5) node [align=center] at (16.5,3) {$+$};

\draw[->, thick] (-3,3) -- (-1,3) node [align=center] at (-3.5,3.5) {$x$};
\draw[->, thick] (21,3) -- (23,3) node [align=center] at (22.3,3.5) {$\hat{y}$};

\node [align=center] at (6, 3) {$\vdots$}; \node [align=center] at (14, 3) {$\vdots$};

\node [align=center] at (10,3) {\large $H$}; \node [align=center] at (8.6,0.5) {\tiny$N_{\rm TX}$};
%\node [align=center] at (8.6,8.5) {\tiny$1$}; \node [align=center] at (8.6,6.5) {\tiny$2$};
 \node [align=center] at (8.6,4.5) {\tiny$1$};

\node [align=center] at (11.4,0.5) {\tiny$N_{\rm RX}$};
%\node [align=center] at (11.4,8.5) {\tiny$1$}; \node [align=center] at (11.4,6.5) {\tiny$2$}; 
\node [align=center] at (11.4,4.5) {\tiny$1$};

\end{tikzpicture}
\caption{Analog beamforming based transmitter (left) and receiver (right) use a bank of phase shifter to perform beamforming in the RF domain. This architecture uses just one pair of A/D or D/A at the baseband.}
\label{fig:txrxABF}
\end{figure}
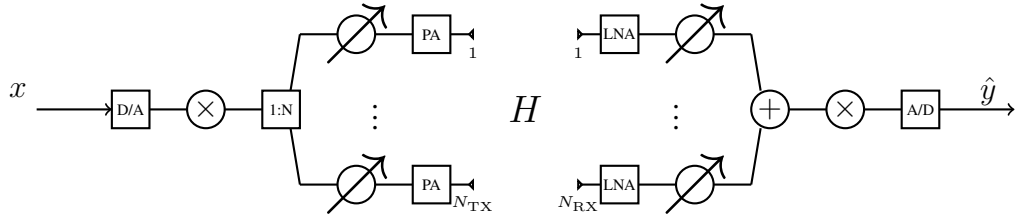

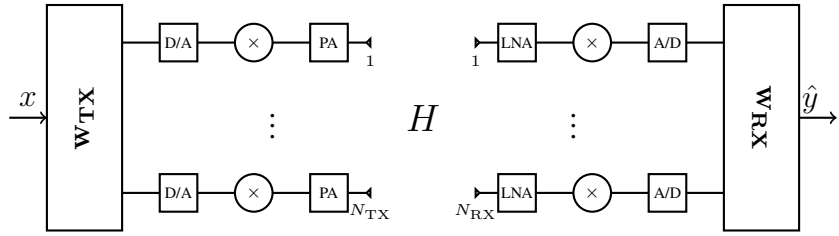
\begin{figure}[ht!]
\centering
\begin{tikzpicture}[scale=0.5]
\draw[thick] (0,0) rectangle (2,6) node [align=center,rotate=90] at (1,3.0) {$\mathbf{w_{\tiny TX}}$};
\draw[thick] (18,0) rectangle (20,6) node [align=center,rotate=270] at (19,3.0) {$\mathbf{w_{\tiny RX}}$};

\foreach \y in {1,5} {
\draw[thick] (2,\y) -- (3,\y); 
\draw[thick] (3,\y-0.5) rectangle (4,\y+0.5) node [align=center] at (3.5, \y) {\tiny D/A};

\draw[thick] (18,\y) -- (17,\y); 
\draw[thick] (17,\y-0.5) rectangle (16,\y+0.5) node [align=center] at (16.5, \y) {\tiny A/D};

\draw[thick] (4,\y) -- (5,\y); \draw[thick] (16,\y) -- (15,\y);
\draw[thick] (5.5,\y) circle (0.5) node [align=center] at (5.5,\y) {\tiny $\times$}; 
\draw[thick] (14.5,\y) circle (0.5) node [align=center] at (14.5,\y) {\tiny $\times$}; 

\draw[thick] (6,\y) -- (7,\y); \draw[thick] (14,\y) -- (13,\y);
\draw[thick] (7,\y-0.5) rectangle (8,\y+0.5) node [align=center] at (7.5, \y) {\tiny PA};
\draw[thick] (13,\y-0.5) rectangle (12,\y+0.5) node [align=center] at (12.5, \y) {\tiny LNA};
\draw[thick] (8,\y) -- (8.5,\y); \draw[thick] (12,\y) -- (11.5,\y);

\draw[thick] (8.5,\y) -- (8.6,\y+0.1) -- (8.6,\y-0.1) --cycle;
\draw[thick] (11.5,\y) -- (11.4,\y+0.1) -- (11.4,\y-0.1) --cycle;
}

\draw[->, thick] (-1,3) -- (0,3) node [align=center] at (-0.5,3.5) {$x$};
\draw[->, thick] (20,3) -- (21,3) node [align=center] at (20.3,3.5) {$\hat{y}$};

\node [align=center] at (6, 3) {$\vdots$}; \node [align=center] at (14, 3) {$\vdots$};

\node [align=center] at (10,3) {\large $H$}; \node [align=center] at (8.6,0.5) {\tiny$N_{\rm TX}$};
%\node [align=center] at (8.6,8.5) {\tiny$1$}; \node [align=center] at (8.6,6.5) {\tiny$2$};
 \node [align=center] at (8.6,4.5) {\tiny$1$};

\node [align=center] at (11.4,0.5) {\tiny$N_{\rm RX}$};
%\node [align=center] at (11.4,8.5) {\tiny$1$}; \node [align=center] at (11.4,6.5) {\tiny$2$}; 
\node [align=center] at (11.4,4.5) {\tiny$1$};

\end{tikzpicture}
\caption{Fully digital transmitter (left) and receiver (right) use a pair of DACs and ADCs per RF stream. Beamforming is performed in the baseband digital domain.}
\label{fig:txrx}
\end{figure}

Most current mmWave designs use \emph{analog} \cite{IBM2017PhasedArray} or \emph{hybrid beamforming} \cite{samsBeamForm}. In these cases, beamforming is performed in radio frequency (RF) or at an intermediate frequency (IF) through a bank of phase shifters (PSs) -- one per antenna element as shown in Fig. \ref{fig:txrxABF}. This architecture reduces the power consumption by using only one pair of analog to digital converters (ADC) and digital to analog converters (DAC) at the Rx and Tx, respectively, per digital stream.  While analog and hybrid beamforming are power efficient, they are only capable of transmitting in one or a few directions at a given time \cite{KhanPi:11-CommMag}. This essentially limits their multiplexing capabilities. In contrast, in \emph{fully digital architectures} \cite{Zhang:05,Heuvel2012,MoHeathTsp15,Jacobsson2017}, shown in Fig. \ref{fig:txrx}, beamforming is performed in baseband. Each RF chain has a pair of ADCs at the Rx and DACs at the Tx enabling the transceiver to simultaneously direct beams in theoretically infinite directions at a given time. 
But, for wide-band systems high precision ADCs and DACs can be very power hungry. To be energy efficient, fully digital beamformers need to use converters with one or few bits of resolution \cite{Singh2009Limits}.

\subsection{Signal Processing with Low Resolution Quantizers}
For communication systems, the degradation due to low resolution converters can be viewed as the introduction of \emph{quantization noise} in the signal. Low resolution converters can be simply viewed as a noise source and the introduction of this additional noise has the effect of lowering the achievable link capacity.
Studies on point-to-point links \cite{Singh2009Limits, MoHeathTsp15, Mollen2017TWC} have demonstrated that operations over wide band channels with low resolution ADCs can achieve sufficient spectral efficiency.  Even with a single bit of precision, as shown in \cite{MoHeathTsp15, Mollen2017TWC}, wide band multi-antenna systems can achieve considerably high spectral efficiency when perfect channel information is assumed.  Further, MIMO channel estimation for wide band systems has been recently studied under the low resolution limit in \cite{Mo2018ChanEst,Mezghani2018ChanEst}.  Moreover, the information theoretic work in \cite{oner2015adc} shows that for point-to-point systems, digital beamforming offers higher rates than analog for a given power budget. A comparison with hybrid beamforming in \cite{moheath2017}, similarly, shows that low resolution digital BF can achieve higher rates than hybrid beamforming while having similar or even lower power consumption.

\subsection{Motivation and Contributions}
The key objectives for 5G cellular systems are achieving higher rates, serving much denser networks, and ultra low latency. The abundant bandwidth in mmWave frequencies could achieve the first two objectives; the third objective, we argue, can be obtained as a consequence of utilizing digital beamforming. 
To make this argument we note that  the fully digital architecture enables greatly enhanced spatial flexibility. 
The works in \cite{barati2015directional,barati2016initial, giordani2018} showed how digital BF can reduce control plane latency by a factor of 10 compared with the alternative, i.e., analog and hybrid BF. Additionally, frequency division multiple access (FDMA) scheduling is feasible with digital BF which enables very efficient transmission of short data and control packets \cite{dutta2017frame}. However, to maintain similar power consumption levels as analog BF, fully digital arrays should operate at low quantization levels. Hence, there is a fundamental trade-off between directional search and spatial multiplexing on the one hand and quantization noise on the other.

The addition of quantization noise does not adversely effect control signaling as these are designed to operate in low SNRs where quantization noise is not dominant \cite{dutta2017frame}. On the other hand, as observed from the theoretical works of \cite{Singh2009Limits, MoHeathTsp15}, the presence of quantization noise essentially limits the maximum achievable rate in the high SNR regime. Thus, in this work we study the effect of low resolution quantization on the \emph{data plane} of mmWave cellular systems. More importantly, we answer the key question as to how many bits of resolution are required, both at the transmitter (DACs) and the receiver (ADCs), to operate in the mmWave bands. Our major contributions in this work are as follows:
\begin{itemize}
\item First, we provide a detailed assessment of various components of a mmWave front-end based on state-of-the-art circuits.  This provides a first order model to analyze the power consumption of different beamforming architectures.
\item Second, we propose an additive quantization noise model (AQNM) to model the signal degradation due to low resolution. We show that our proposed model accurately predicts system behavior both at low SNR and high SNR regimes.
\item Third, we show that for most practical cellular operations the effect of low resolution quantization on the achieved rate is negligible. In fact, we show that at the Rx, $3-4$ bits of resolution is sufficient for wide band mmWave applications.
\item Finally, we study the effects of low resolution DACs on mmWave transmissions. We show that with no assumption on additional filtering, $4$-bits of resolution is sufficient to guarantee 3GPP new radio (NR) compliant adjacent carrier leakage. Further, analyzing the error vector magnitude (EVM) we show that 4-bits of DAC resolution is sufficient both at BS and UE.
\end{itemize}

The rest of this paper is organized as follows. We model the power consumption of the mmWave front end circuit in Sec. \ref{Sec2:circts}. In Sec. \ref{Sec:2NktModel}, we describe the downlink (DL) system model for mmWave systems and the available multiple access schemes. We present the proposed AQNM in Sec. \ref{Sec:AQNM} and discuss on the effects quantization noise under practical operating conditions. Next, in Sec. \ref{Sec5}, we present the model of a DAC from a signal processing point of view and detail the effects of low resolution quantization on the transmitted signal quality. In Sec \ref{sec6} we validate the proposed AQNM and study its effect on system rate. Further, using extensive simulation we also determine the resolution needed for the DACs at the mmWave Tx. Sec. \ref{SecConc} concludes the paper. This work was presented in part in \cite{dutta17Asilomar}.

%Before we begin our analysis of fully digital transceivers at mmWave, in this section we make a comprehensive survey of the current literature on mmWave radio circuits. 
\section{Power Consumption in mmWave Radio Circuits} \label{Sec2:circts}
%The recent work \cite{yang2018} is, to the best of our knowledge, the first to demonstrate a system which performs fully digital BF at mmWave frequencies. This system uses very high precision DACs and ADCs and achieves very high spectral efficiency. 
Power efficient fully digital beamformers will have to rely on low resolution converters DACs and ADCs. To better understand the effect of decreasing the precision of the converters, in this section we model the power consumption of transceiver front ends (FE) at mmWave frequencies for analog, hybrid and fully digital beamforming. 

\paragraph*{RF Front End} The RF front end (RFFE) refers to the circuitry between the antenna and the baseband data converters (DACs or ADCs). As shown in Fig.\ref{fig:txrxABF} and Fig.\ref{fig:txrx}, this includes the power amplifiers (PAs) or low noise amplifiers (LNAs), mixers, PSs, combiners and splitters.
%From Fig.\ref{fig:txrxABF} and Fig.\ref{fig:txrx}, the RF front end (RFEE) is the circuit from the input of the mixer to the output of the power amplifier (PA) at the Tx, and from the input to the low noise amplifier (LNA) to the generated baseband signal at the output of the mixer at the Rx. 
At the Tx, consider that the total power delivered by the base-band circuit is $P_{\rm in}^{\rm BB}$. The mixers, splitters and PSs are considered to be passive devices which introduce insertion loss (IL) but do not draw any power.
From Fig. \ref{fig:txrxABF} it is easy to see that for analog and $K$-stream hybrid beamformer, the power of input signal at the PA is given as
\beq
P_{\rm in,ana}^{\rm PA} = P_{\rm in, hyb}^{\rm PA}= P_{\rm in}^{\rm BB} - 10\log_{10}(N_{\rm Tx}) -  IL_{\rm PS}  - IL_{\rm mix} ~(\text{dBm}),
\eeq
where $IL_{\rm PS}$ and $IL_{\rm mix}$ is the IL due to the PS and mixer respectively, and $10\log_{10}(N_{\rm Tx})$ is the loss in signal power due to the $1:N_{\rm Tx}$ power splitter. 
%For a transmitter with $N_{\rm Tx} = 16$ and with a total radiated power of $23$~dBm, $P_{\rm RF} = 12.58$~mW. Further, when $ IL_{\rm PS} = 10$~dB, we can get, $ P_{\rm DC}^{PA} = 50.2 ~\text{mW}$, for each antenna element when $\eta_{\rm PA} = 25\%$.
Similarly, from Fig. \ref{fig:txrx}, for the fully digital beamformer, we can write,
\beq
P_{\rm in, dig}^{\rm PA} = P_{\rm in}^{\rm BB} - 10\log_{10}(N_{\rm Tx}) - IL_{\rm mix}~(\text{dBm}).
\eeq
To transmit an output power $P_{\rm out}^{\rm RF}$~dBm, the D.C power drawn by the PA is
\beq
P_{\rm DC,BF}^{\rm PA} = \frac{1}{\eta_{\rm PAE}} \left(10^{0.1 P_{\rm out}^{\rm RF}} -  10^{0.1 P_{\rm in,BF}^{\rm PA}}\right)~\text{~mW},
\eeq
where $\eta_{\rm PAE}$ is the power added efficiency of the PA and $P_{\rm in,BF}$ is the input power given a beamforming (BF) architecture. Note that for a given effective isotropic radiated power (EIRP), for an $N_{\rm Tx}$ antenna system, $P_{\rm out}^{\rm RF} = {\rm EIRP} - 20\log (N_{\rm Tx})$. The total power drawn by the Tx RFFE can hence be given as
\beq 
P_{\rm Tx}^{\rm RFFE} = N_{\rm Tx} P_{\rm DC,ana}^{PA} + N_s P_{\rm LO},
\label{eq:2-4}
\eeq 
where $N_s$ is the number of baseband streams; $N_s = 1$ for analog BF, $N_s = K$ for $K$-stream hybrid BF and $N_s = N_{\rm Tx}$ for fully digital BF.
Based on (\ref{eq:2-4}), the power consumption of the Tx RFFE is reported in Table \ref{tab:powerTx} where the calculations are based considering ${\rm EIRP} = 30$~dBm, $IL_{\rm PS} = 10$~dB, $IL_{\rm mix} = 6$~dB, $P_{\rm LO} = 10$~dBm and $\eta_{\rm PAE} = 20\%$.

At the Rx, LNAs are characterized by their figure of merit (FoM) which relates the gain $(G_{\rm LNA})$ and the noise figure $(N_{\rm LNA})$ to the D.C. power drawn $(P_{\rm dc}^{\rm LNA})$ as \cite{Song2008}
\beq
P_{\rm dc}^{\rm LNA} = \frac{G_{\rm LNA}}{{\rm FoM}( N_{\rm LNA}-1)},
\eeq
in linear scale. The total RFFE power consumption at the Rx is thus,
\beq 
P_{\rm Rx}^{\rm RFFE} = N_{\rm Rx} P_{\rm dc}^{\rm LNA} + N_s P_{\rm LO},
\eeq 
where $N_s$ is the number of baseband streams at the Rx. Now, for the digital BF, if the LNA gain is selected as $G_{\rm LNA,dig}$, then for analog/hybrid BF the required LNA gain required will be $(G_{\rm LNA,dig} + IL_{\rm PS})$ which compensates for the IL due to the RF PSs. In Table. \ref{tab:powerTx}, we list the Rx RFFE power consumption given $G_{\rm LNA, dig} = 10$~dB, $IL_{\rm PS} = 10$~dB, $N_{\rm LNA} = 3$~dB and a LNA FoM $= 6.5 ~\text{mW}^{-1}$ for $90$~nm CMOS as reported in \cite{Adabi07}.

\paragraph*{Gain Control at the Rx}
% Given the LNA gain ($G_{\rm LNA}$) the signal power at the output of the mixers for hybrid BF would be 
% \beq
% P_{\rm out, hyb}^{\rm mix}(d) = P_{\rm Rx} (d) - 10\log(K) + G_{\rm LNA} - IL_{\rm PS} + 10\log(N_{\rm Rx}) - IL_{\rm mix},
% \label{eq:2-8}
% \eeq
% where, $P_{\rm Rx}(d) = {\rm EIRP_{\rm Tx}} - {\rm PL}(d)$ where ${\rm PL}(d)$ is the path loss at a Tx-Rx separation of $d$~m. With $K = 1$ in (\ref{eq:2-8}) we get the expression for the analog beamformer. On the other hand, for digital beamforming we have,
% $ P_{\rm out, dig}^{\rm mix}(d) = P_{\rm Rx} + G_{\rm LNA} - IL_{\rm mix}.$
Given a fixed Tx EIRP, the power received at the Rx is $P_{\rm Rx}(d) = {\rm EIRP_{\rm Tx}} - {\rm PL}(d),$ where ${\rm PL}(d)$ is the path loss for a Tx-Rx separation of $d$~m.
To maintain a constant baseband power of $P_{\rm BB}^{\rm out}$, the variable gain amplifier (VGA) at the input of the ADC needs a gain range from $0 - G_{\rm max}^{VGA}$~dB. Noting that $G_{\rm LNA}$ is adjusted to compensate for $IL_{\rm PS}$ for analog/hybrid BF, to drive a total baseband power of $P_{\rm BB}^{\rm out}$, the VGA gain range required is
\beq 
G_{\rm max}^{VGA} = P_{\rm BB}^{\rm out} - 10\log (N_{\rm Rx}) + IL_{\rm mix} - (G_{\rm LNA} - IL_{\rm PS}) - P_{\rm Rx} (d = d_{\rm cell}),
\eeq 
where $d_{\rm cell}$ is the radius of the cell.
For a down-link (DL) transmission with ${\rm EIRP_{\rm Tx}} = 43$~dBm; at the cell edge, $d_{\rm cell} = 100$~m, $P_{\rm Rx}(d = d_{\rm cell}) = -87$ dBm for a mmWave non-line of sight channel \cite{AkdenizCapacity:14}. Assuming similar values of IL as on the Tx RFFE and considering $G_{\rm LNA, dig} = (G_{\rm LNA} - IL_{\rm PS}) = 10$~dB, to maintain $P_{\rm BB}^{\rm out} = 10$~dBm we require a gain range of $G_{\rm max}^{VGA} = 82$~dB.

The figure of merit of a VGA  $(\rm FoM_{\rm VGA})$ is defined by \cite{wang2012} as
\beq
\rm FoM_{\rm VGA} = \frac{G_{\rm max}^{\rm VGA} \times f_{\rm BW} }{\rm P_{\rm dc}^{\rm VGA} \times {\rm A_{\rm chip}}},
\eeq
where $G_{\rm max}^{\rm VGA}$ is in dB, the bandwidth $f_{\rm BW}$ is in GHz, the D.C power draw $P_{\rm dc}^{\rm VGA}$ in mW and the VGA active area $A_{\rm chip}$ is in mm$^2$. The FoM reported by \cite{wang2012} for a $90$-nm CMOS process with an active area of $0.01 \text{mm}^2$ is $5280$. Considering the same active area, we report the power drawn by the VGA(s) for the three beamforming architectures in Table \ref{tab:powerTx}.

\paragraph*{DAC and ADC} For wide band wireless applications the data converters, DACs and the ADCs, are considered to be the most power hungry elements. The power consumed by an ADC or a DAC ($P_{\rm conv}$) is a linear function of the sampling frequency $(f_s)$ and grows exponentially with the number of bits of resolution $(n)$ as
\begin{equation}
P_{\rm conv} = \text{FoM} \times f_s \times 2^n,
\label{eq:PowerConv}
\end{equation}
where $\text{FoM}$ is the figure of merit of the converter.  As mmWave systems are envisioned for ultra wide-band applications, the sampling frequencies are in the order of $1$~GHz.
In analog or hybrid beamforming the use of one or a few pairs of converters limit the power consumption. For fully digital systems, a reduction of $n$ is hence the only way to reduce the power consumption. 

Contrary to the assumption made in \cite{alkhateeb2014mimo}, we observe that both DACs and the ADCs are equally power hungry. For instance the 4-bit Flash based ADC designed in \cite{Nasri2017} has a $\textrm{FoM} = 65$~fJ/conv, while a state of the art DAC proposed in \cite{Olieman2015DAC} has a $\textrm{FoM} = 67.6$~fJ/conv. Thus, a pair of $8$-bit ADC consumes $33.28$~mW of power at $f_s = 1$~GHz. At the same sampling rate, a pair of $8$-bit DAC consumes $34.6$~mW of power, nearly same as that of the ADC. Similar trends can be observed in more recent works \cite{Juanda2018, Kim2016DAC}. Hence, using low resolution DACs can considerably reduce the power consumption by fully digital Tx as shown in Table \ref{tab:powerTx}.

\paragraph*{Filtering at the Tx} The output of the DACs will require analog low pass filters (LPF) to reject spectral images, and maintain out of band emission limits as discussed in Sec. \ref{Sec5}. In this work we assume the use of active switched capacitor filters. For an $m$-th order active LPF with cutoff frequency at $f_c$, the power per pole per Hertz FoM is given as \cite{Houfaf2012},
\beq
\text{FoM}_{\rm LPF} = {P_{\rm dc}^{\rm LPF}}/{(m \times f_c)}.
\eeq
For wide band LPFs, based on \cite{Houfaf2012, Chen2013}, we consider the $\text{FoM} = 1.3 $~mW/GHz. For mmWave beamformers, as discussed in Sec. \ref{sec:6c}, we can use a first order LPF with $f_c = 400$~MHz, each of these filters will thus consume a power of $P_{\rm dc}^{ \rm LPF} = 0.52$~mW. Depending on the BF architecture, the total power drawn by the LPF is equal to $N_s P_{\rm dc}^{ \rm LPF}$, with $N_s = 1$ for analog BF.

% \begin{align}
% P_{\rm ana,Tx}^ {\rm tot} &= P_{\rm ana}^{\rm RF,Tx} + P_{\rm DAC Subsys.} (n), \\
% P_{\rm dig,Tx}^ {\rm tot} &= P_{\rm ana}^{\rm RF,Tx} + N_{\rm Tx} P_{\rm DAC Subsys.} (n) \\
% P_{\rm ana,Rx}^ {\rm tot} &= P_{\rm ana}^{\rm RF,Rx} + P_{\rm ADC Subsys.} (n), \\
% P_{\rm dig,Rx}^ {\rm tot} &= P_{\rm ana}^{\rm RF,Rx} + N_{\rm Rx} P_{\rm DAC Subsys.} (n).
% \end{align}

\begin{table*} [!t]
\centering
\renewcommand{\arraystretch}{0.8}
\begin{tabular}{|| p{4cm}|p{2cm}|p{1cm}|p{2cm}|p{2cm}|p{1cm} ||} \hline  \hline 
 \multicolumn{6}{||l||}{\bf Rx Front End Power Consumption [mW]} \\ \hline 
BF Arch.			& RFFE 	        &  VGA      & ADC~($8$ bits) & ADC~($4$ bits) & Total  \\\hline
Analog		 	    & 257.3 	    &  1.55     & 33.3  	     & --             & 292.15 \\ \hline
Hybrid ($K=2$) 	    & 267.3 	    &  3.11     & 66.6  	     & --             & 337.01 \\ \hline
Digital (High res.) & 184.7 	    &  24.85    & 532.8	     & --                 & 742.35 \\ \hline
Digital (Low res.) 	& 184.7  	    &  24.85    & -- 			 & 33.3 	      & 242.85 \\ \hline  \hline 
 \multicolumn{6}{||l||}{\bf Tx Front End Power Consumption [mW]} \\  \hline 
BF Arch.			& RFFE 	        &  LPF      & DAC ($8$ bits) & DAC ($4$ bits)  & Total  \\\hline
Analog		 	    & 321.2 	    &  0.52     & 34.4	         & --     	       & 356.12 \\ \hline
Hybrid ($K=2$) 	    & 331.2 	    &  1.04     & 69.2 	         & --              & 401.44 \\ \hline
Digital (High res.) & 459.9	        &  8.32     & 553.6          & --              & 1021.82 \\ \hline
Digital (Low res.) 	& 459.9	        &  8.32     & -- 	         & 34.4 	       & 502.62 \\ \hline  \hline 
\end{tabular}
\caption{Power consumption (in mW) for each component in the RF chain for various receiver and transmitter architectures with 16 Tx and Rx antennas.}
\label{tab:powerTx}
\end{table*}

From Table \ref{tab:powerTx}, we see that both at the Tx and the Rx, low resolution quantizers can considerably reduce the power consumption of the front-end circuitry. At the Rx, the use of low resolution quantizers for digital BF front ends leads to  a power draw lower than even the analog BF. This reduction in power comes at the cost of increased quantization noise in the system. We will analyze the effect of coarse quantization in the sequel.

\section{mmWave Downlink System Model} \label{Sec:2NktModel}
In this section we detail the network in which mmWave cellular transceivers operate. To study the effect of low quantization converters, we characterize the signal to interference and noise ratio (SINR) under different multiple access strategies.

\subsection{Network Model}
We consider a wireless network with $N$ base stations (BS) each with $N_{\rm BS}^{\rm ant}$ antennas. Each BS serves a multiplicity of UEs each with $N_{\rm UE}^{\rm ant}$ antennas. The BSs and UEs operate over mmWave frequencies i.e., $f_c >10$ GHz, where $f_c$ is the carrier frequency. Downlink (DL) and uplink (UL) transmissions use the same channel using time division duplexing (TDD).
To mitigate the high isotropic path loss at mmWave frequencies, the BSs and UEs employ digital beamforming with low resolution DACs and ADCs both at the Tx and the Rx.
%Power and chip area constraints are more relaxed at the BS than at hand-held UEs. Therefore, for our analysis in this section, we assume that DACs used at the BS have higher resolution than the ADCs at the UEs. This simplifies our analysis as the quantization noise induced by the BS DAC becomes negligible compared to the low resolution ADC at the UE. A more general case is explored in Sec. \ref{sec6} where both Tx and Rx have same quantizer resolution.
%For example, if the resolution of ADC at the UEs is $b_0$ bits while at the BSs the DAC has $b_0+2$ bits of resolution, the quantization noise introduced by the BS DAC is nearly 6 dB lower than the UE ADC for DL reception.

\paragraph{DL Transmission}
In this system, a BS transmits single streams of data to $K = |\mathcal{K}(t)|$ associated UEs, where $K\leq \min(K_0, N_{\rm BS}^{\rm beam})$, $\mathcal{K}(t)$ is the set of scheduled users at time instant $t$, $K_0$ is the total number of UEs associated with the BS, and $N_{\rm BS}^{\rm beam}$ is the number of beams supported by the BS. Under a fixed power budget of $P$ watts, the transmitted signal is given by
\begin{equation}
\mathbf{x}(t) = \sum_{k \in \mathcal{K}(t)} \rho\mathbf{v}_{k} s_k(t),
\end{equation}
where $\rho = \sqrt{P/K}$, $\mathbf{v}_{k} \in \mathbb{C}^{N_{\rm BS}^{\rm ant}\times 1}$ is the transmit side long-term beamforming vector between user $k$ and the BS. Throughout this work we assume long-term beamforming \cite{AkdenizCapacity:14} where BF vectors are computed based on the channel covariance matrices. Without loss of generality, we assume $P=1$ and throughout this work we considered transmit power is split equally among all beams by the BS. At each time instant $t$ the BS schedules $\mathcal{K}(t)$ UEs which can be multiplexed in time, frequency or spatially. We discuss the relevant multiple access schemes in Section \ref{secMA}.

\paragraph{DL Reception} The signal received at the $k$-th UE, \emph{before} digital beamforiming is applied, is given as
\begin{equation}
\mathbf{y}_k(t) = \rho\mathbf{H}_k\mathbf{v}_{k} s_k(t) + \sum_{j \in \mathcal{K}(t),j\neq k} \rho\mathbf{H}_k\mathbf{v}_{j} s_j(t) + \mathbf{z}_k + \mathbf{n}_k,
\label{eq:2-2}
\end{equation}
where $\mathbf{H}_k \in \mathbb{C}^{N_{\rm UE}^{\rm ant}\times N_{\rm BS}^{\rm ant}}$ represents the channel matrix between user $k$ and the BS, $\mathbf{z}_k \in \mathbb{C}^{N_{\rm UE}^{\rm ant}\times 1}$ represents the inter-cell interference and $\mathbf{n}_k \in \mathbb{C}^{N_{\rm UE}^{\rm ant}\times 1}$ represents the receiver noise. Note that the second term on the right hand side of (\ref{eq:2-2}) accounts for the \emph{intra-cell interference} (ICI). The receiver noise is assumed to be zero mean i.i.d. Gaussian with covariance matrix given as $\sigma_n^2 I_{N_{\rm UE}^{\rm ant}}$. Similarly, the inter-cell interference $\mathbf{z}_k$ is also be assumed to be i.i.d Gaussian  with the covariance matrix $\sigma_z^2 I_{N_{\rm UE}^{\rm ant}}$.

\subsection{DL Multiple Access}\label{secMA}

MmWave BSs employing fully digital beamforming are capable of transmitting to multiple directions at the same time. Unlike conventional analog/hybrid architectures this allows the multiplexing of large number of UEs, spatially or in frequency, leading to potential increase in system throughput, and enabling low latency transmissions. The multiple access techniques available for fully digital BS transmissions are as follows.

\paragraph{TDMA} Time division multiple access (TDMA) based DL will have $|\mathcal{K}(t)| = 1$, i.e., only one UE can be scheduled for transmission at any given time instance and will have access to the entire bandwidth. As user allocations are orthogonal in time, the ICI is zero. Yet, TDMA can potentially lead to wastage of allocated bandwidth especially for UEs with bursty or low rate traffic. Sophisticated scheduling and frame design, such as \cite{dutta2017frame}, is required to intelligently exploit the large bandwidth available at mmWave.

\paragraph{OFDMA} A key attraction of digital beamforming at mmWave frequencies is that it enables orthogonal frequency division multiple access (OFDMA). In OFDMA, BS allocates chunks of total bandwidth i.e., physical resource blocks (PRBs) to multiple UEs at each scheduling instance. The scheduler allocates a variable portion of the bandwidth $W_k \geq 0$ to $k = 1,2\ldots K_0$ UEs based on their data requirement, channel condition,  and scheduling priority.
Due to the use of orthogonal channels for transmission, OFDMA based DL signals do not encounter ICI. Moreover, unlike TDMA, multiple users can be schedules at each time slot leading to higher utilization of mmWave bands and faster transmission of small packets. This is especially attractive for low latency communication.

\paragraph{SDMA} Fully digital beamforming has the potential to support space division multiple access (SDMA) as the BS Tx can transmit to multiple users on separate spatial streams\footnote{Although the point-to-point mmWave link is low rank, additional degrees of freedom can be achieved by multi-user MIMO}. This has the potential of increasing the available degrees of freedom $K$ folds, where $K$ is the number of streams. In this case $|\mathcal{K}(t)| = K(t)$, the ``optimal'' number of streams that can be supported at time $t$. A key challenge for SDMA based transmission is mitigating the effect of the ICI. To minimize ICI, the BS scheduler must carefully select users, or beamforming patterns, or both, thus limiting the number of beams over which transmissions occur at any scheduling instance.

\section{Link Layer AQNM Model} \label{Sec:AQNM}
\begin{figure}
\centering
\begin{tikzpicture}[scale=1]
\draw[->] (4,1) -- (4.5,1) node [align=center] at (4,1.15) {$x$};
\draw[thick] (4.5,0.5) rectangle (5.5,1.5) node [align=center] at (5,1) {$1-\alpha$};
\draw[->] (5.5,1) -- (6,1);
\draw[thick] (6.25,1) circle (0.25) node [align=center] at (6.25,1) {$+$};
\draw[->] (6.25,2) -- (6.25,1.25) node [align=center] at (7,2.25) {\small $v_q \sim \mathcal{CN}(0,\sigma_v^2)$};
\draw[->] (6.5,1) -- (7,1) node [align=center] at (7,1.25) {$x_q$} ;
\end{tikzpicture}
\caption{Additive quantization noise model for low resolution quantizer. The parameter $\alpha$ is the inverse coding gain and models the quantizer resolution (e.g., $\alpha=0$ implies infinite resolution).}
\label{fig:aqnm}
\end{figure}
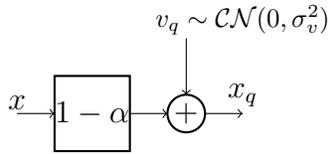

\subsection{Effective SINR}
 We first derive an analytical model for the effective SINR under the finite quantization limit for a multi-antenna receiver. For this purpose, we use a slightly modified version of the additive quantization noise model (AQNM) as presented in \cite{FletcherRGR:07}. In our AQNM, shown in Fig. \ref{fig:aqnm}, the effect of finite uniform quantization of a scalar input $y$ is represented as a constant gain plus an additive white Gaussian noise (AWGN). Furthermore, \cite{FletcherRGR:07} showed that if an input complex sample $y$ is modeled as a random variable, then the quantizer output $y_q$ can be written as
\begin{equation} \label{eq:Qmodel} 
	y_q = Q(y) = (1-\alpha)y + v_q, \quad \Exp|v_q|^2 = \alpha(1-\alpha)\Exp|y|^2,
\end{equation}
where $Q(\cdot)$ denotes the quantization operation and $v_q$ represents quantization errors uncorrelated with $y$ and approximated as a complex Gaussian. The parameter $\alpha \in [0,1]$ is the \emph{inverse coding gain} of the quantizer, i.e., $\alpha=0$ implies infinite quantizer resolution. The parameter $\alpha$ is assumed to depend on the resolution of the quantizer and is independent of the input distribution.

%In the following section we analyze the effect of this impairment induced by finite quantization on the received signal.

%\section{Reception with low resolution quantizers} \label{sec4}
We now extend our model to a multi-antenna receiver model. For the received signal in (\ref{eq:2-2}), each component $y_k^i(t)$ of the received signal $\mathbf{y}_k(t)$ is independently quantized by an ADC before an appropriate receiver-side beamforming vector $\wbf_{\rm UE}$ is applied. Thus, from \eqref{eq:2-2} and \eqref{eq:Qmodel}, the quantized received vector is given as
\begin{align}
\mathbf{\hat{y}}_k (t) & = \Qbf (\mathbf{y}_k(t)) \nonumber \\
& = (1 - \alpha) \left( \rho\mathbf{H}_k\mathbf{v}_{k} s_k(t) +  \sum_{j \in \mathcal{K}(t), j\neq k} \rho\mathbf{H}_k\mathbf{v}_{j} s_j(t)  +  \mathbf{z}_k + \mathbf{n}_k \right) + \vbf.
\label{eq:Qsig}
\end{align}
The vector $\vbf$ denotes the additive quantization noise with covariance 
$\sigma_v^2 \Ibf_{N_{\rm UE}^{\rm ant}}$. We assume that the quantization errors
across antennas are uncorrelated. From \eqref{eq:Qsig}, the average per component energy to the input $\ybf$ of the quantizer is
\begin{align}
	\frac{1}{N_{\rm UE}}\Exp\|\ybf\|^2 &= 
    \rho^2 E_k + \rho^2\sum_{j \in \mathcal{K}(t),j\neq k} E_j + \sigma_{\zbf}^2 + \sigma_n^2,  
\end{align}
where $E_j = (1/N_{\rm UE}^{\rm ant})\Exp\|\Hbf_k \vbf_{j}s_j\|^2~\forall j$ is the average
received symbol energy per antenna for each stream $j\in \mathcal{K}$. From \eqref{eq:Qmodel}, the quantization noise variance is 
\begin{equation} \label{eq:sigvq}
	\sigma_v^2 = \alpha(1-\alpha)\left[ \rho^2 E_k + \rho^2 \sum_{j \in \mathcal{K}(t), j\neq k} E_j + \sigma_{z}^2 + \sigma_n^2 \right].
\end{equation}
Note that for (\ref{eq:sigvq}) we use the fact that at any given $t$ the inter-cell interference is independent of the transmitted signal, i.e., $\Exp\left[z_k^H (\Hbf_k \vbf_j s_j) \right] = 0,~ \forall j \in \mathcal{K}(t),$
and symbols transmitted to different users are independent, i.e. $\Exp\left[s_i^*s_j \right] = 0$, for point to point links, hence
\beq
\Exp\left[\left(\Hbf_k \vbf_k s_k \right)^H \Hbf_j \vbf_j s_j  \right] = 0; \quad \forall j \in \mathcal{K}(t), j \neq k.
\eeq 
%Additionally we note that for broadcast channels $\Exp\left[s_i^*s_j \right] \neq 0$, and the quantization noise power increases due to channel correlation. For cellular systems such a scenario is of little importance. %since broadcast messages require very low bit rates and can be transmitted via OFDMA to all users.

After applying a receiver-side beamforming vector $\ubf_{\rm k}$, the channel between the UE and the BS is an effective SISO channel. Define the Rx side BF gain on signal in stream $j \in \mathcal{K}(t)$  as
\beq \label{eq:BFGains}
	G_j := \Exp|\ubf_{\rm k}^*\Hbf_k\vbf_{\rm j} s_j|^2 / E_j,
\eeq
which is the ratio of the signal energy after beamforming to the received signal 
energy per antenna. We note that in (\ref{eq:BFGains}) if transmit beamforming vectors are chosen such that $\Hbf_k\vbf_{\rm j} = \mathbf{0},~ \forall j\neq k, j\in \mathcal{K}(t)$, then the ICI at user $k$ is zero, but this requires careful beam planning and scheduling and may not, except for channel conditions, be even possible. On the other hand, to remove ICI one can set $s_j = 0, \forall j\neq k, j\in \mathcal{K}$, which is achieved by orthogonal transmissions schemes (TDMA and OFDMA). Although the latter approach for mitigating ICI simplifies scheduling, there is a loss in the available degrees of freedom.

Observe that, if there is no quantization error $(\alpha=0)$, the beamformed SINR of user $k$ is
\begin{align} \label{eq:gamBF}
	\gamma_k^{\rm BF}&:= \frac{\rho^2\Exp|\ubf_{k}^H\Hbf_k\vbf_{k} s_k|^2}{
    	\sigma^2_n + \sigma^2_z + \rho^2\sum_{j\neq k}\Exp|\ubf_{k}^*\Hbf_j\vbf_{j} s_j|^2 }  = \frac{\rho^2 G_k E_k}{\sigma^2_n + \sigma^2_z + \rho^2\sum_{j\neq k} G_j E_j} = \frac{\gamma_k'}{1+ \sum_{j\neq k} \gamma_j'},
\end{align}
where $\gamma_j' = \rho^2 G_j E_j/(\sigma_n^2 + \sigma_z^2), \forall j \in \mathcal{K}$. Now, with finite resolution quantization, using the AQNM, the signal after beamforming is given by
%Note that to obtain a MIMO channel we may simply replace the beamforming vectors $\wbf_{\rm BS}$ and $\wbf_{\rm UE}$ with unitary matrices where each column is an eigenvector of the channel matrix $H$.
\begin{align}\label{eq:QsigBF}
 y_k^{\rm Q,BF} := \ubf_k^H \hat{\ybf}_k 
     = & ~\rho (1 - \alpha) \ubf_k^H \Hbf_k \vbf_k s_k \nonumber \\  
     & + \rho (1 - \alpha) \sum_{j\neq k} \ubf_k^H\Hbf_k \vbf_j s_j + (1 - \alpha) \ubf_{k}^H \zbf + (1 - \alpha) \ubf_k^H \nbf + 
    \ubf_k^H \vbf. 
\end{align} 
Without loss of generality, assuming $\|\ubf_{k}\| = 1$ the mean beamformed received signal energy is
\begin{equation} \label{eq:Esq}
	E_k^{\rm BF} = (1-\alpha)^2\rho^2\Exp|\ubf_{\rm k}^H\Hbf_k\vbf_{\rm k}s_k|^2 = 
    \rho^2(1-\alpha)^2 G_{\rm k}E_k,
\end{equation}
while the average noise plus interference energy is
\begin{equation} \label{eq:Enq}
	W_{ k}^{BF} = (1-\alpha)^2\left(\sigma^2_z + \sigma^2_n + \rho^2 \sum_{j\neq k} G_j E_j\right) + \sigma^2_v.
\end{equation}
Finally, combining \eqref{eq:sigvq}, \eqref{eq:gamBF},
\eqref{eq:Esq}, and \eqref{eq:Enq}, we obtain an expression for the SINR after beamforming as
\begin{equation} \label{eq:gamQBF}
	\gamma^{\rm Q,BF}_k = \frac{E_k^{\rm BF}}{W_{k}^{\rm BF}} = \frac{(1-\alpha)\gamma_{k}'}{1 + (1-\alpha)\sum\limits_{j\neq k}^{}\gamma_{j}' + \alpha\left(\frac{\gamma_{k}'}{G_{k}}+  \sum\limits_{j\neq k}^{}\frac{\gamma_{j}'}{G_j}\right)}.
\end{equation}

\subsection{Orthogonal Transmission}\label{sec:4b}
For orthogonal DL transmissions, i.e., TDMA or OFDMA, (\ref{eq:gamQBF}) simplifies as
\begin{equation} \label{eq:gamO-QBF}
	\gamma^{\rm Q,BF}_k = \frac{(1-\alpha)\gamma_{k}^{\rm BF}}{1 +  (\alpha/{G_{k}})\gamma_{k}^{\rm BF}},
\end{equation}
where we use the fact that $\gamma_k^{\rm BF} = \gamma_k'$ in the absence of ICI. Using  \eqref{eq:gamO-QBF}, we can qualitatively understand the system-level effects of quantization by looking at the following two regimes: 
\paragraph{Low SNR} In the low-SNR (or SNR) regime, $\gamma_k^{\rm BF}$ is small, hence
\beq
	\gamma^{\rm Q, BF}_k \approx (1-\alpha)\gamma_k^{\rm BF},
\eeq
i.e., the SINR is decreased only by a factor $1-\alpha$. We show in Sec. \ref{sec6} that at moderate quantization levels this has very little impact on system performance.
\paragraph{High SNR} In this regime as $\gamma_k^{\rm BF} \arr \infty$
\beq \label{eq:OrthoSaturation}
	\gamma^{\rm Q,BF}_k \arr \frac{G_{k}(1-\alpha)}{\alpha}.
\eeq
Thus, the effect of quantization is to saturate the maximum SINR i.e., the effect of finite quantization is critical only at high SNR (or SINR). In Sec. \ref{sec6} we show that even for $3-4$ bits of resolution, the effects caused at the high SINR limit are not significant for cellular systems.

\subsection{SDMA Transmission} \label{sec:sdmaAnalysis}
Comparing (\ref{eq:gamQBF}) with (\ref{eq:gamO-QBF}) we note that for SDMA transmission the effect of quantization noise is further enhanced by the presence of ICI. In (\ref{eq:gamQBF}), we note that
\beq \label{eqGainSimplify}
G_j = \frac{\Exp\|\ubf_{\rm k}^H\Hbf_k\vbf_{\rm j} s_j\|^2}{E_j} =  N_{\rm UE}^{\rm ant} \frac{\Exp \|\ubf_{\rm k}^H\Hbf_k\vbf_{\rm j} s_j \|^2}{\Exp\|\Hbf_k \vbf_{j}s_j\|^2}.
\eeq
Under the assumption that $\| s_j\|^2 = 1$, using the Kronecker model \cite{kermoal2002stochastic} we obtain
\[
\Exp \|\ubf_{\rm k}^*\Hbf_k\vbf_{\rm j} s_j \|^2 = \frac{1}{N_{\rm UE}^{\rm ant}} \left(\ubf_k^H \Qbf_k^{\rm rx}  \ubf_k \right) \left(\vbf_j^H \Qbf_k^{\rm tx}  \vbf_j \right),
\]
where $\Qbf_k^{\rm rx} = \Exp|\Hbf_k  \Hbf_k^H|$ and $\Qbf_k^{\rm tx} = \Exp|\Hbf_k^H  \Hbf_k|$ are the receive and transmit covariance matrices respectively for the channel between the BS and user $k$. Similarly,
\beq \label{eq:txbf}
\Exp\|\Hbf_k \vbf_{j}s_j\|^2 = \left(\vbf_j^H \Qbf_k^{\rm tx}  \vbf_j \right).
\eeq
\noindent Thus we can rewrite (\ref{eqGainSimplify}) as
\beq \label{eqGainSimplify2}
G_j = \ubf_k^H \Qbf_k^{\rm rx}  \ubf_k = G_k, ~\forall j\in \mathcal{K},
\eeq
which points simply to the fact that for any signal received by the $k$-th UE from the associated BS will have the same receiver-side beamforming gain.

The presence of ICI in SDMA systems essentially limits the maximum achievable SINR. From (\ref{eq:txbf}), we note that although ICI can be eliminated by selecting transmit beamforming vectors such that
$\left(\vbf_j^H \Qbf_k^{\rm tx}  \vbf_j \right) \to 0, ~\forall j\neq k,$
but this can be a very hard problem in practice and in some cases a solution may not exist. Thus, in our analysis we will assume that ICI is always present for SDMA based systems.

For this, we simplify (\ref{eq:gamBF}) by representing the ICI term as
\beq \label{eq:gammaknj}
\sum_{j\neq k} \gamma_j' = \psi \gamma_k', \quad \psi \geq 0,
\eeq
where $1/\psi$ is the signal to interference ratio (SIR), and as $\gamma_k' \to \infty$, $\gamma_k^{\rm BF} \to (1/\psi)$. An interesting observation for SDMA systems is that even in the absence of quantization noise the SINR saturates depending on the severity of the ICI.

\paragraph*{Effect of low resolution quantization} Using (\ref{eqGainSimplify2}), and (\ref{eq:gammaknj}) in (\ref{eq:gamQBF}) we can express the SINR in the presence of quantization noise and ICI as
\beq \label{eq:sinrICI1}
{\gamma_k^{\rm Q,BF} } = \frac{(1-\alpha)\gamma_k'}{1 + (1-\alpha) \psi\gamma_k' + (\psi+1)\frac{\alpha}{G_k}\gamma_k'}.
\eeq 
Furthermore, as in  (\ref{eq:gamO-QBF}) we can rewrite (\ref{eq:sinrICI1}) as
\beq 
{\gamma_k^{\rm Q,BF} } = (1-\alpha \beta){\gamma_k^{\rm BF} },
\eeq
where
\beq \label{eq:betaDef}
\beta = \frac{1 + (\psi+1)\gamma_k'/G_k}{1 + (1-\alpha)\psi\gamma_k' + \alpha(\psi+1)\gamma_k'/G_k}.
\eeq 
Finally, from (\ref{eq:betaDef}) we note that  $\beta < \frac{1}{\alpha}$ is always satisfied under the AQN model (note that $\beta > \frac{1}{\alpha} \Leftrightarrow \alpha > 1$, which is not admissible). It follows from (\ref{eq:betaDef}) that for $\beta < 1$ we must have
$G_k > 1 + \frac{1}{\psi}$, and hence,
\beq \label{eq:GkIneq}
{\gamma_k^{\rm Q,BF} } > (1-\alpha){\gamma_k^{\rm BF} }, \text{~if~} G_k > 1 + \frac{1}{\psi}.
\eeq
We note that when $\psi$ is \emph{large}, i.e., when ICI dominates, it is easy to satisfy the inequality in $(\ref{eq:GkIneq})$. SINR degradation in this regime is dominated by the large ICI and \emph{not} quantization noise. Alternatively, as $\psi \to 0$ the system converges to the orthogonal transmission case discussed in the previous subsection.

To summarize our analysis in this section, for multi-user communications with long term beamforming and low resolution front ends, we state the following:
\begin{itemize}
\item For orthogonal transmission (i.e. FDMA or TDMA) in low SNR/SINR regime there is very little or effectively no loss due to low resolution quantization. 
\item For orthogonal transmission in high SNR or SINR regimes there exists a saturation of the effective SNR or SINR due to quantizer resolution.
\item For transmission schemes when orthogonality within the cell is not always guaranteed (e.g., SDMA), the degradation in SINR is dominated by the ICI.
\end{itemize}
In Sec. \ref{sec6} we will validate these claims using extensive simulations.

% \section{DL User Scheduling}
% some text maybe.

\section{Signal Impairments with Low Resolution DAC} \label{Sec5}
The previous section analyzed quantization noise at the Rx. In this section, we analyze the Tx side. Low resolution quantization at the Tx can result in quantization noise being present both in band and out-of-band.  
For cellular communication systems, 3GPP specifies signal characteristics that must be adhered to by any transmitter. The key characteristics are (a) output power, (b) the adjacent carrier leakage ratio (ACLR), and (c) transmitted signal quality specified by the error vector magnitude (EVM).
The output power is not affected by the addition of quantization noise but the EVM of the Tx increases due to the addition of this noise source. Moreover, the quantization noise is not band limited as opposed to the signal and contributes to an increase in the power leaked into the adjacent carrier. This out-of-band distortion is rather constraining for cellular communications systems and hence 3GPP specifies ACLR limits that all commercial transmitters must comply with. In this section we show the effects of low resolution DACs on the transmitted signal, and in the process determine the precision required by the DACs in the mmWave transmitter to meet 3GPP regulations.

\subsection{Model of the DAC}\label{Sec5a}
\begin{figure}[!t]
\centering
\begin{tikzpicture}
\draw[->] (-1,0.5) -- (0,0.5) node [align=center] at (-1,0.75) {$x_{\rm BB}$}; 
\draw[thick] (0,0) rectangle (1,1) node [align=center] at (0.5,0.5) { $\mathbf{\uparrow m}$} node [align=center] at (0.5,1.5) {Interpolate} ;
\draw[->] (1,0.5) -- (2,0.5); 
\draw[thick] (2,0) rectangle (4,1)  node [align=center] at (3,0.5) { $\mathbf{ Q(\cdot)}$};
\draw[->] (4,0.5) -- (5,0.5) node [align=center] at (4.5,0.75) {$x_{q}$}; 
\draw[thick] (5,0) rectangle (7,1) node [align=center] at (6,0.5) {\bf ZOH};
\draw[->] (7,0.5) -- (8,0.5);
\draw[thick] (8,0) rectangle (10,1) node [align=center] at (9,0.5) {\bf LPF};
\draw[->] (10,0.5) -- (11,0.5) node [align=center] at (11,0.75) {$x_{a}$};
\draw[dashed] (1.75, -0.25) rectangle (7.25, 1.25) node [align=center] at (4.5 , 1.5) {DAC};
\end{tikzpicture}
\caption{High level model of a digital to analog converter at baseband. The DAC is clocked at $f_s$ where $f_s = m f_{\rm BB}$.}
\label{fig:DacModel}
\end{figure}
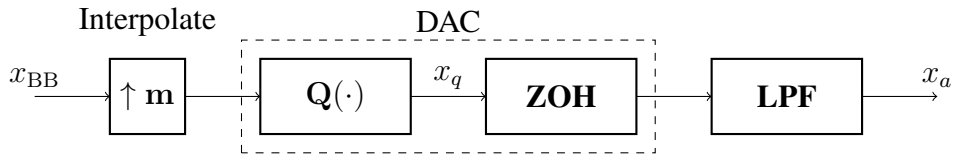

To model the effect of low resolution on the transmit signal, we first describe a model for the DAC. A DAC comprises of a quantizer $Q(\cdot)$ and the zero order hold (ZOH) circuit. An output analog low pass filter (LPF), as shown in Fig. \ref{fig:DacModel}, is used to attenuate the spectral images of the signal located at an interval of $f_s$, where $f_s$ is the sampling rate of the DAC.
The sampling rate must be at least  $f_s = f_{\rm BW}$, where  $f_{\rm BW}$ is the bandwidth of the input signal $x_{\rm BB}$. In most designs, like \cite{Alavi2014}, $f_s = m\times f_{\rm BW}$, with $m > 1$ so that the spectral images formed at the output of the DAC are spaced sufficiently apart in frequency. Thus the base band signal is interpolated, i.e., upsampled by $m$ and filtered, before it is converted to analog. 
The interpolator not only relaxes the design of the analog LPF but also minimizes the distortion caused by the ZOH on the in-band component. 
Additionally, upsampling the signal by a factor of $m$ also reduces the power of the quantization noise by the same factor, e.g., a $m = 2$ interpolation of the baseband signal will lead to lowering the quantization noise by $3$~dB. 
This gain comes at the cost of doubling the sampling rate $f_s$ of the DAC which, from (\ref{eq:PowerConv}), doubles the power consumed. As pointed out in \cite{Alavi2014}, high over sampling is not practical for wide band systems due to the linear increase in power consumption.

\subsection{Adjacent Carrier Leakage} \label{Sec5b}
From our analysis in Section \ref{Sec:AQNM}, quantization noise for low resolution converters can be modeled as white Gaussian noise. This implies that the quantization noise has a flat spectrum while the signal of interest is band limited. This is problematic for practical systems as this noise causes unwanted interference in the adjacent bands.
For interoperability, cellular transmissions need to limit the  amount of power leaked into the adjacent bands. This restriction is quantified, by 3GPP, as the \emph{adjacent carrier leakage ratio} (ACLR) defined as,
\beq 
{\rm ALCR} = 10 \log_{10} \left(\frac{P_{\rm in}}{P_{\rm ac}} \right),
\eeq 
where $P_{\rm in}$ is the total power in the transmission channel and $P_{\rm ac}$ is the total power accumulated over a given adjacent channel.

Well known effective techniques, like windowed overlap and add (WOLA) OFDM \cite{Wola}, are used to reduce the ACLR in LTE systems. These techniques focus on reducing the inter modulation products and operate on the signal in the digital domain. Quantization noise due to finite resolution DACs, however, is introduced in these signals after digital processing. Thus techniques like WOLA have no effect on the quantization noise.
A classical method of dealing with quantization noise is the use of $\Delta\Sigma$ feedback structures. Recent works on Rx beamforming like \cite{PalgunaLowRes16} have considered such circuits for low resolution receivers to ``clean up'' the in band signal. Although attractive, as pointed out in \cite{MehrpooJssc18,deltaSigmaISSCC17}, such techniques rely on circuits that eliminate matching and timing errors, which increases the power consumption considerably. Moreover, $\Delta\Sigma$ modulators also require the DAC to operate with a high oversampling ratio, which further increase the power consumption.

Thus, the only practical option to control the ACL due to quantization noise is by imposing stricter restrictions on the analog LPF which, on the first glance, implies that higher order filters will be required when low resolution DACs are used at the transmitter. Thus the filters will either take more space on the chip (when they are passive) or consume higher power (for active CMOS filters). In Sec. \ref{sec:6c} we perform extensive simulations to determine the order of the LPF that meets the ACLR requirements at the Tx. More importantly, in the sequel we show that, for moderately low resolution DACs, no additional restrictions are imposed on the analog LPF.  

\subsection{Transmitted Signal Quality} \label{sec:EvmTh}
As in  \cite{3GPPTS38104}, we quantify the transmitted signal quality by its EVM. Intuitively, the EVM captures the error in the modulated symbol produced due to Tx impairments. It is considered a key factor in determining the maximum modulation order a transmitter can faithfully support. The EVM, $\epsilon$, for a Tx signal is given as
\beq 
\epsilon^2 = \frac{\Exp_{t,f} |Z(t,f) - I(t,f) |^2 }{\Exp_{t,f} |I(t,f) |^2},
\label{eq:evm}
\eeq 
where $Z(t,f)$ and $I(t,f)$ are the received symbol and ideal modulation symbol respectively at time $t$ and sub-carrier $f$. From (\ref{eq:evm}) it is clear that lower the value of $\epsilon$ the cleaner is the transmitted signal and $\epsilon$ must be small in order to support high order modulations as they are more sensitive to distortions.

The signal impairments introduced by the mmWave RFFE, including the local oscillator (LO) phase noise, LO leakage, I-Q imbalance, etc. can be modeled as an AWGN noise source following the work in \cite{Gupta2012}. We represent the RF impairments as  zero mean complex Gaussian random variable $n_{\rm RF} \sim \mathcal{CN}(0, \sigma_{\rm RF}^2)$. Based on the AQNM in (\ref{eq:Qmodel}), we can rewrite (\ref{eq:evm}) as
\beq 
\epsilon^2 = \alpha^2 + \frac{\sigma_{\rm RF}^2 + \sigma_{\rm v}^2}{\Exp |I(t,f) |^2},
\eeq 
where $\sigma_{\rm v}^2$ is the variance of the quantization noise. Thus the presence of quantization noise effectively limits the EVM from going to 0 even when $\sigma_{\rm RF} \to 0 $. Thus, low resolution quantizers essentially limit the maximum spectral efficiency that can be achieved by limiting the highest modulation order that can be supported by a transmitter. This is crucial for the utilization of the large bandwidths available at mmWave frequencies.

In Sec. \ref{sec:6c}, we perform extensive simulations to analyze the effect of low resolution DACs on transmitted signal quality. We show, in the sequel, that low resolution DACs can be used for mmWave transmitters under 3GPP specified limits on ACLR and EVM.

\section{Simulations Results}\label{sec6}
In this section we present our results obtained through link level and cellular simulations. Firstly, we verify the AQNM presented in Section \ref{Sec:AQNM} through a series of link layer OFDM simulations. Next, we use a multi-cell multi-user simulation at 28 GHz to study the effect of low resolution quantization and multiple access schemes on link quality and throughput. Finally, we investigate the effect of low resolution DACs on the transmitted signals.

\subsection{Verification of the AQNM}
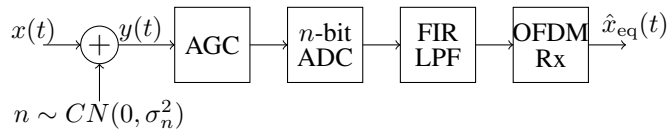
\begin{figure}
\centering
\begin{tikzpicture} [scale=0.5]
\draw (1,1) circle (0.5) node [align=center] at (1,1) {$+$};
\draw [->] (1.5,1) -- (3,1) node [align=center] at (2.1,1.4) {\small $y(t)$}; \draw[->] (-.5,1) -- (0.5,1) node [align=center] at (-0.75,1.35) {\small $x(t)$}; \draw [->] (1,-0.5) -- (1,0.5) node [align=center] at (1,-0.85) {\small $n \sim CN(0,\sigma_n^2)$};
\draw (3,0) rectangle (5,2) node [align=center] at (4,1) {\small AGC};
\draw (6,0) rectangle (8,2) node [align=center,font=\small\linespread{0.8}\selectfont] at (7,1)  {\small $n$-bit \\ \small ADC};
\draw (9,0) rectangle (11,2) node [align=center,font=\small\linespread{0.8}\selectfont] at (10,1)  {\small FIR \\ \small LPF};
\draw [->] (5,1) -- (6,1) ; \draw [->] (8,1) -- (9,1) ; \draw [->] (11,1) -- (12,1) ;
\draw (12,0) rectangle (14,2) node [align=center,font=\small\linespread{0.8}\selectfont] at (13,1)  {\small OFDM \\  \small Rx};
\draw [->] (14,1) -- (15,1) node [align=center] at (15.25, 1.5) {\small $\hat{x}_{\rm eq} (t)$};
\end{tikzpicture}
\caption{Simulation model for Rx front end with low resolution ADC modeled as a $n$-bit scalar quantizer.}
\label{fig:rxBlocks}
\end{figure}

To verify the proposed AQNM for low resolution converters, we use a link level OFDM simulator. The simulation parameters, given in Table \ref{tab:aqnSimPar}, are from the 3GPP NR standards \cite{3GPPTS38211}. We consider a wide band AWGN channel with a receiver structure shown in Fig. \ref{fig:rxBlocks}. 
\begin{table}
\begin{center}
\renewcommand{\arraystretch}{0.8}
\begin{tabular}{|>{\raggedright}p{6 cm}|>{\raggedright}p{4 cm}|}
  \hline
  {\bf Parameter} & {\bf Value}
  \tabularnewline \hline
Channel bandwidth $(f_{\rm BW}^{\rm ch})$ & 400 MHz
\tabularnewline \hline

FFT size $(N_{\rm fft})$ & $4096$
\tabularnewline \hline

Subcarrier spacing  & $120$~kHz
\tabularnewline \hline

OFDM chip rate $(f_{\rm chip})$ & $491.52$ MHz
\tabularnewline \hline

Subcarriers per PRB  & $12$
\tabularnewline \hline

Max. PRBs used $(N_{\rm PRB}^{\rm max})$ & $275$
\tabularnewline \hline

% Max. occupied bandwidth, $W$ & 396 MHz
% \tabularnewline \hline

Symbol duration &  $10.67 \mu$s.
\tabularnewline \hline
\end{tabular}
\caption{OFDM parameters for link level simulations.}
\label{tab:aqnSimPar}
\end{center}
\end{table}
The ADC is modeled as a finite resolution scalar quantizer. The automatic gain control (AGC) ensures that the input to the quantizer has a unit variance. A digital finite impulse response (FIR) LPF is used at the output of the ADC to remove out of band noise. Here, we note that practical transceivers will also employ analog LPFs before the ADC or the AGC to eliminate adjacent bands. As we do not model adjacent carrier blocking, in this simulation we omit the analog filtering.

\begin{figure}[!t]
\centering
\subfloat [$n_{\rm ADC} = n; ~ n_{\rm DAC} = n + 2$; $N_{\rm PRB} = 274$] {
	\includegraphics[width=0.45\textwidth]{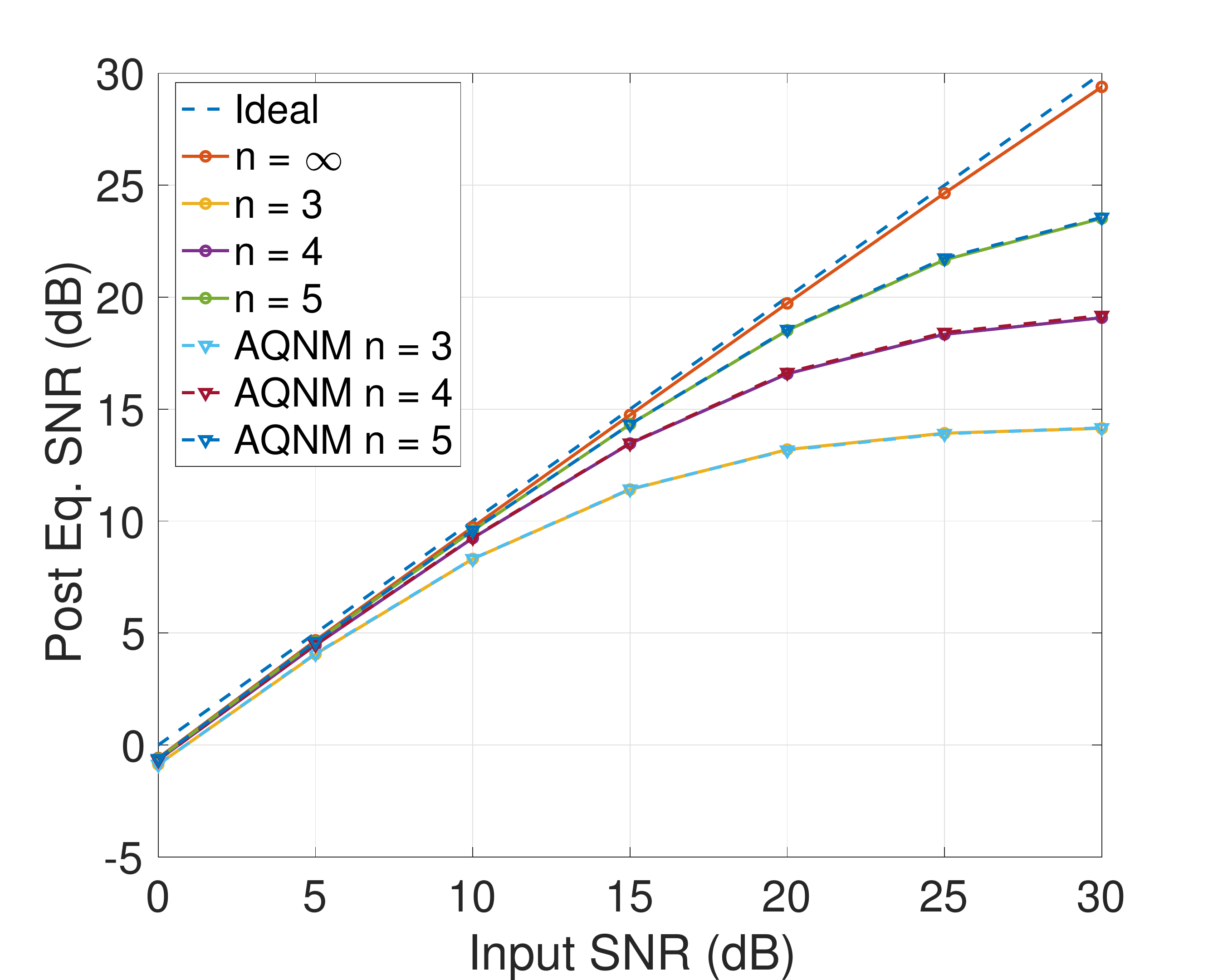} \label{fig:aqn1a}
} 

\subfloat [$n_{\rm ADC} = n_{\rm DAC} = n$; $N_{\rm PRB} = 274$] {
	\includegraphics[width=0.45\textwidth,height=0.35\textwidth]{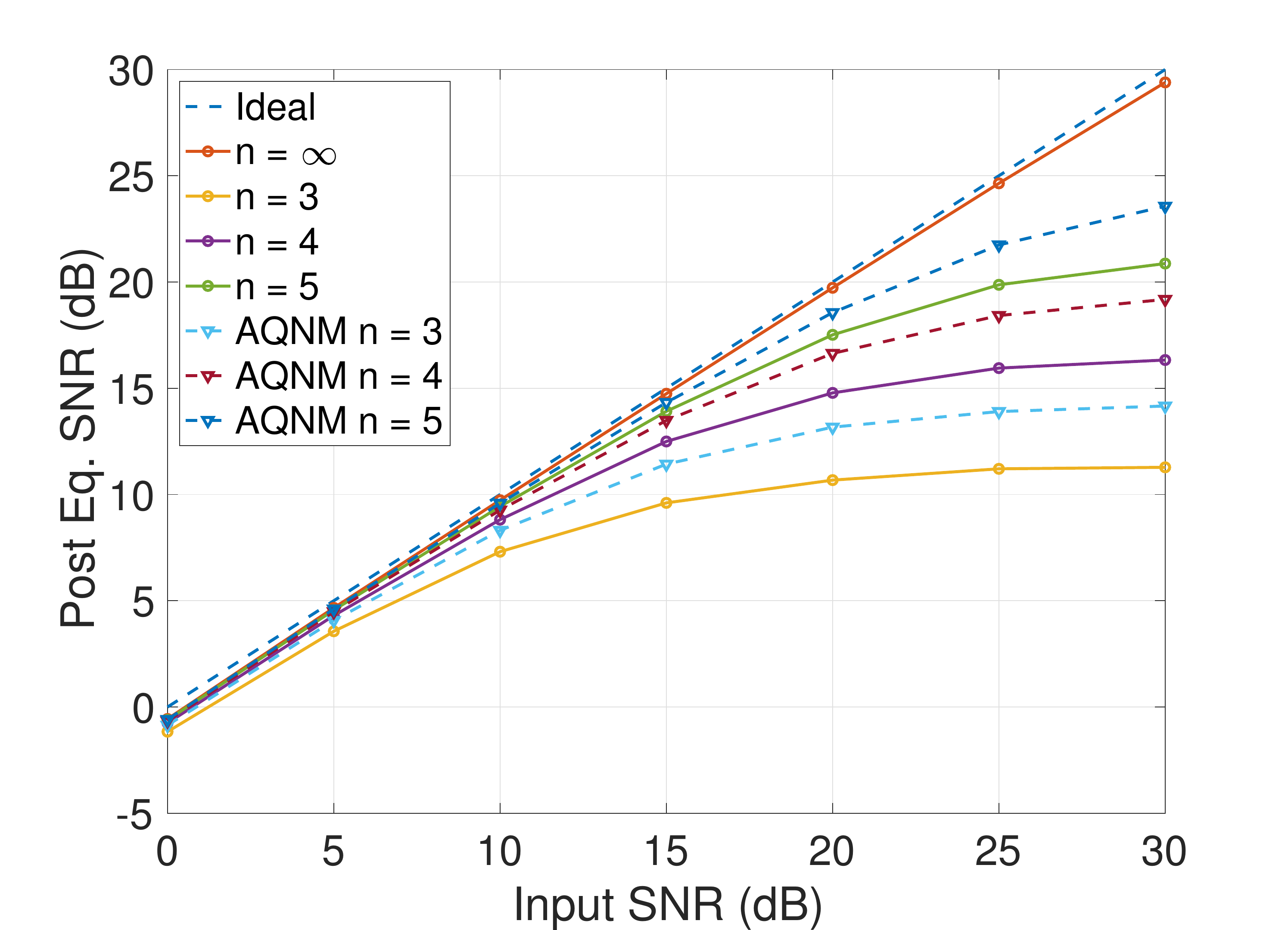} \label{fig:aqn1b}
} \qquad
\subfloat [$n_{\rm ADC} = n_{\rm DAC} = n$; $N_{\rm PRB} = 200$] {
	\includegraphics[width=0.45\textwidth,height=0.35\textwidth]{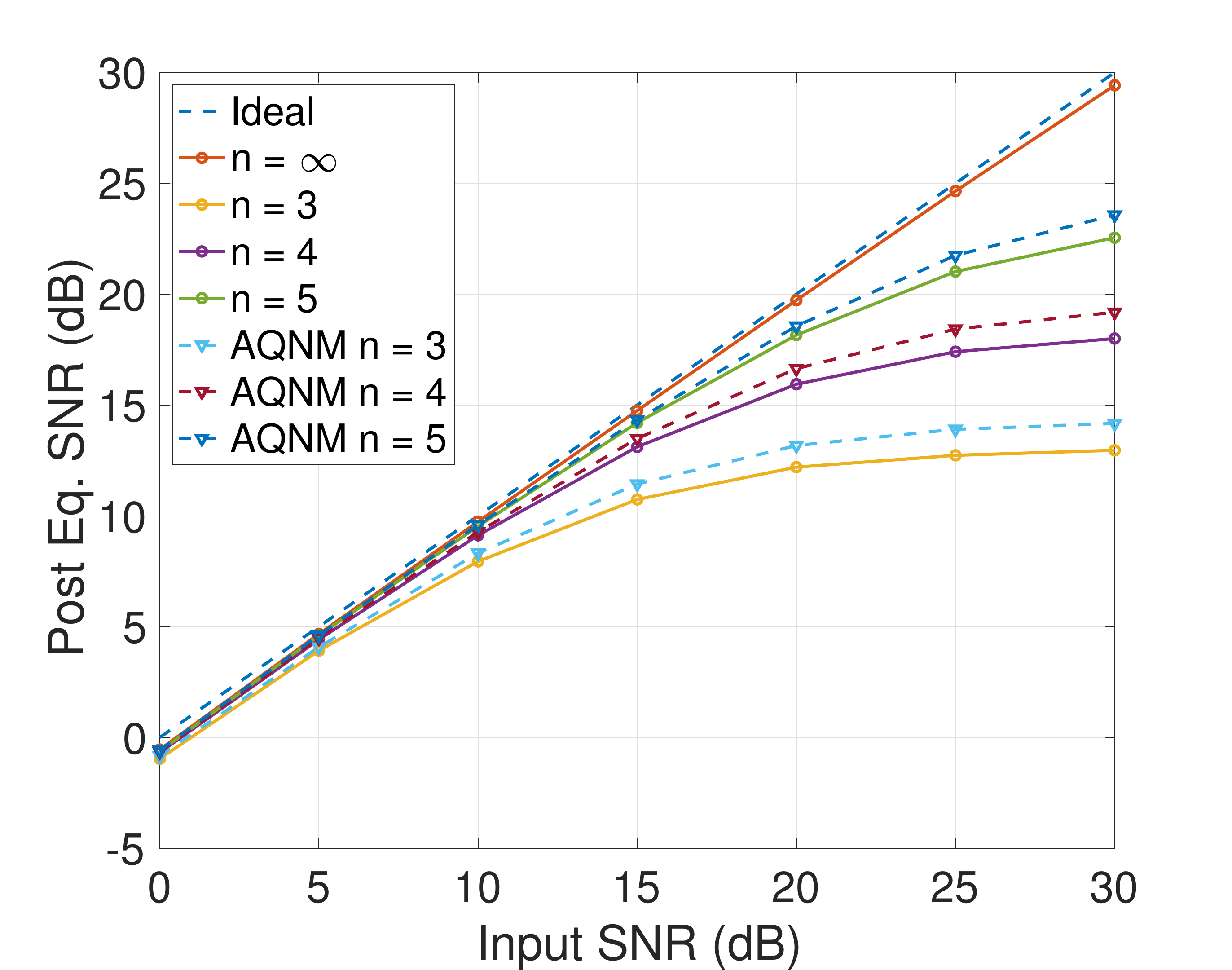} \label{fig:aqn1c}
}
\caption{Post-equalization SNR as a function of input SNR for varying quantization levels. Fig. (\ref{fig:aqn1a}) shows the accuracy our proposed AQNM with link level OFDM simulation based on 3GPP NR specification under assumption of orthogonal transmission, i.e., no ICI. Fig. (\ref{fig:aqn1b}) and  (\ref{fig:aqn1c}) show the effect of oversampling on quantization noise to highlight the trade-off between quantization noise and spectral efficiency.}
\label{fig:aqnmOrtho}
\end{figure}

\paragraph*{Orthogonal Transmission} Fig. \ref{fig:aqnmOrtho} compares the effective SNR predicted by the AQN model with the simulated post-equalization SNR, for varying quantization levels ($n$). The value of $\alpha$ is computed assuming an optimal uniform $n$-bit quantizer \cite{oner2015adc}. From Fig. (\ref{fig:aqn1a}), we see that the AQNM very accurately predicts the signal degradation due to finite quantizer resolution. Here we consider that the Tx uses a $n+2$-bit DAC, and hence, quantization noise added by the Tx is 6 dB lower than that at the Rx.  We observe that finite quantization  has the effect of saturating the effective SNR in the high SNR regime. On the other hand, with $n > 3$, at SNRs below $15$ dB, the effect of quantization noise on the system is negligible. In the case when both the DAC at the Tx and the ADC at the Rx have $n$-bits of resolution, the quantization noise power doubles. This is observed in Fig. \ref{fig:aqn1b}, where in the high SNR regime, the simulated curve is nearly $3$dB lower that predicted by (\ref{eq:gamO-QBF}). 

More interestingly, comparing Fig. \ref{fig:aqn1b} and Fig. \ref{fig:aqn1c}, we observe the effect of \emph{oversampling} on quantization noise. OFDM systems generally have a OFDM chip rate slightly higher than the signal bandwidth. For instance in 3GPP NR, for a 400 MHz channel the OFDM chip rate is 491.52 MHz. 
%Moreover, 3GPP NR standards for 5G allow smaller chunks of bandwidths, called \emph{bandwidth parts}, to be used for transmission. 
In the high SNR regime, in the presence of quantization noise, for a system with $n$-bit quantizers at both Tx and Rx, we can express the SNR as,
\begin{align}
\rm{SNR} ~(\rm lin.) & = \frac{P_{\rm sig}}{N_0 + \frac {2}{\rm OSR} \sigma_q^2},\nonumber \\
\rm{SNR} ~({\rm dB}) & \approx P_{\rm sig} ({\rm dB}) + 3 + 10\log_{10} (\sigma_q^2) + 10\log_{10} \left({N_{\rm fft}}/{N_{\rm sc}} \right),
\end{align}
where ${\rm OSR} = \frac{N_{\rm fft}}{N_{\rm sc}}$ is the oversampling ratio. The effect of oversampling is shown in Fig. \ref{fig:aqn1b} and Fig. \ref{fig:aqn1c} where by using only 200 PRBs, as opposed to 274, we increase the OSR from $0.95$ dB to $2.32$dB. This points to an interesting trade-off. When the system employs low resolution quantizers, under good channel conditions, oversampling by using a smaller part of the bandwidth can reduce the quantization noise at the cost of spectral efficiency, but without an increase in power consumption. %\footnote{We still run the DAC/ADC at the same sampling rate. Running the DACs and ADCs at higher sampling rate decreases the quantization noise but at the cost of power consumption.}.
In fact, in the high SNR regime, reducing the size of the bandwidth part can enable the use of high modulation and coding schemes boosting rates.

\paragraph*{SDMA} Next, we turn our attention to the effect of quantization noise on systems with in-band ICI. Such scenarios are interesting especially when large arrays are available at the Tx and multiple users can be scheduled on multiple transmit beams simultaneously with all the available bandwidth (SDMA). In practical scenarios, transmitting on multiple orthogonal beams may not be possible due to the nature of the multi-user channel. In such cases, ICI becomes dominant in the system.
\begin{figure}[!t]
\centering
\includegraphics[scale=0.22]{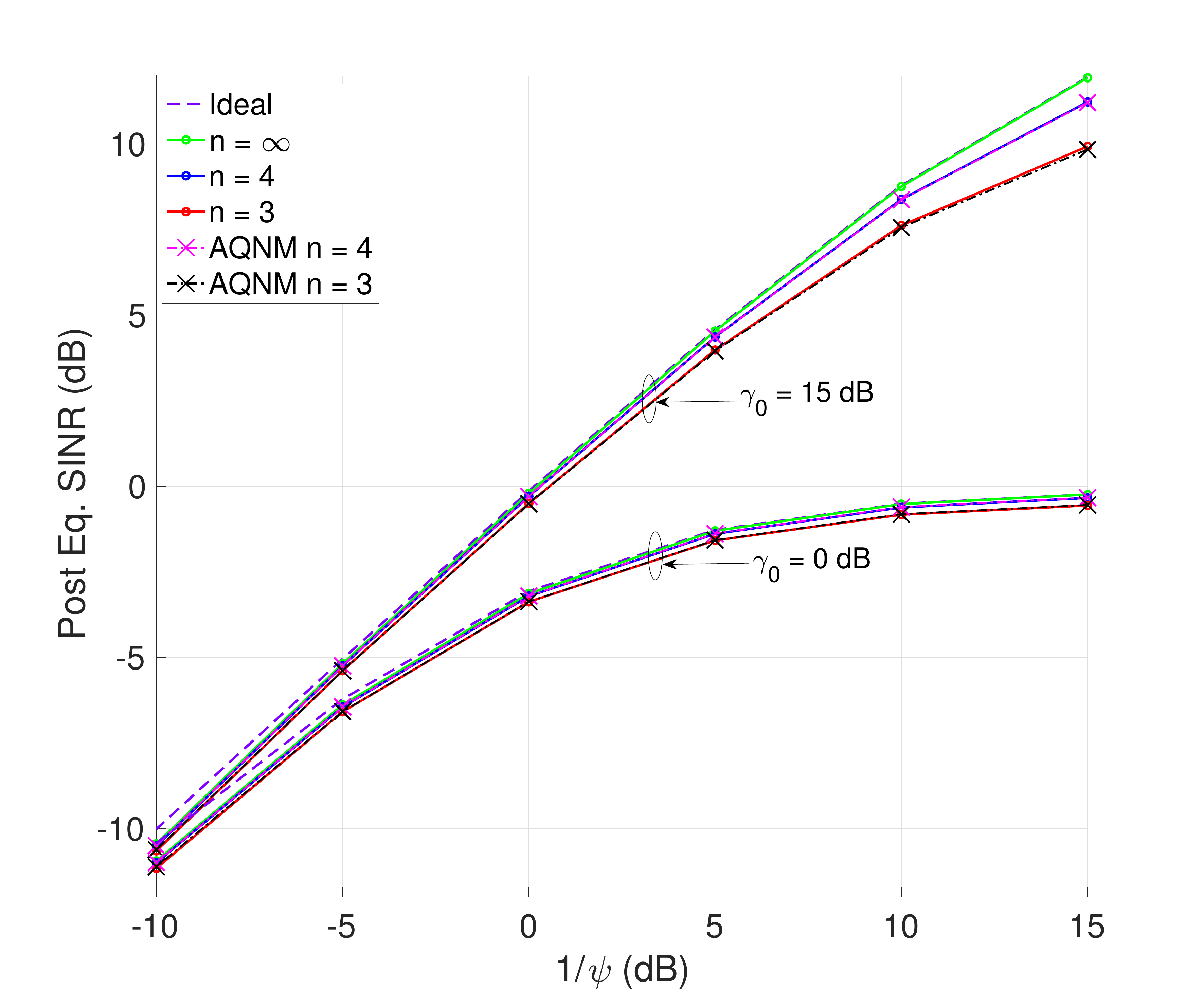}
\caption{Post-equalization SINR as a function SIR ($1/\psi$) with varying quantization levels. Results validate the proposed AQNM using link level OFDM simulation based on 3GPP NR specification in the presence of ICI.}
\label{fig:aqnmSDMA}
\end{figure}
In Fig. \ref{fig:aqnmSDMA} we plot the post-equalization SNR versus the SIR ($1/\psi$). As predicted by (\ref{eq:GkIneq}), when the interference is low (high SIR), without any assumption on processing gains ($G_k$) at the receiver, there is effectively no degradation due to quantization noise. The effect of quantization only becomes perceivable when SIR is high as discussed in Sec. \ref{sec:sdmaAnalysis}. As shown in Fig. \ref{fig:aqnmSDMA}, in noise-limited scenarios ($\gamma_0 = 0$~dB), even at high SIR, the degradation due to quantization noise is less than $0.5$ dB. Substantial loss in link quality is only observed when both SIR and SNR ($\gamma_0$) are high, as seen in Fig. \ref{fig:aqnmSDMA} for $\gamma_0 = 15$~dB. For ADCs with $3$-bits of resolution we observe approximately $2$ dB of loss in the SINR due to finite quantization. For 4-bits of resolution, this loss is less than $1$~dB.

% may be some more text here!

\subsection{Multi-cell Multi-user Simulations} \label{Sec:DLSim}
We apply our link layer AQNM to understand the effect of low-resolution quantization on the DL system capacity. We simulate a 1Km by 1Km area covered by hexagonal cells of radius $100$~m. Each cell is assumed to serve on average $10$~UEs which are randomly ``dropped''. We then compute a random path loss between the BS and the UEs based on the urban mmWave channel model presented in \cite{AkdenizCapacity:14}. We simulate a DL transmission scenario where BSs transmit a single stream to every user. Both BSs and UEs are assumed to perform longterm digital beamforming \cite{Lozano:07} making use of the spatial second-order statistics of channel. For our simulations, we assume that the BSs always have data to send to every UE (full buffer assumption). The relevant parameters for our simulations are summarized in Table \ref{tab:cellSimPar}. 

\begin{table}
\begin{center}
\renewcommand{\arraystretch}{0.8}
\begin{tabular}{|>{\raggedright}p{4 cm}|>{\raggedright}p{4.5 cm}|}
  \hline
  {\bf Parameter} & {\bf Value}
  \tabularnewline \hline

% BS layout and sectorization $W$ & Hexagonal cells on a $2$km by $2$km area with three sectors per cell.
% \tabularnewline \hline

% UE layout & Dropped uniformly with average of 10 UEs per sector 
% \tabularnewline \hline

Cell radius & $100$~m
\tabularnewline \hline

Carrier frequency & $28$~GHz
\tabularnewline \hline

Pathloss model & \cite{AkdenizCapacity:14}
\tabularnewline \hline

DL bandwidth $(W_{\rm tot})$ & $1$~GHz
\tabularnewline \hline

DL Tx power & $35$~dBm
\tabularnewline \hline

Rx noise figure & $8$~dB
\tabularnewline \hline

Max. spectral efficiency & $7.4063$ b/s/Hz
\tabularnewline \hline

BS antenna array & $8 \times 8$ uniform planar
\tabularnewline \hline

UE antenna array & $4 \times 4$ uniform planar
\tabularnewline \hline

BF mode & Digital long-term, single stream
\tabularnewline \hline

Transmission time interval& 125$\mu$s
\tabularnewline \hline

Control overhead & 20\%
\tabularnewline \hline

Traffic Model & Full buffer
\tabularnewline \hline

\end{tabular}
\caption{Multi-cell simulation parameters.}
\label{tab:cellSimPar}
\end{center}
\end{table}

\paragraph*{OFDMA} 
%We demonstrate orthogonal transmission using the OFDMA scheme. In fact, we note that with analog or $K$-stream hybrid beamforming based transmitters, OFDMA cannot be used as a multiple access scheme. 
For OFDMA based cellular systems, within one transmission time interval (TTI), each UE is assigned a non-overlapping part of the total bandwidth by the associated BS. Each link gets full beamforming gain but only uses a part of the total bandwidth. Orthogonalization in frequency eliminates ICI but also limits the maximum achievable rate. To study such a system in a practical setting, we employ a proportional fair scheduling algorithm for medium access control. At the $T$-th TTI, the $k$-th UE associated to BS $j$ is assigned a weight,
\beq 
w_{j,k}^T = \frac{\rho_{j,k}^T}{\sum_{t=0}^{T} r_t} \label{eq:PFwt}
\eeq
where $r_t$ is the data that had been scheduled at the $t$-th TTI, and $\rho_{j,k}^T$ is the spectral efficiency of the link at the $T$-th TTI. Note that, for our simulations we assume that the BS does not have any information about the quantization noise at the Rx. The weights are normalized as $ w_{j,k}^T = {w_{j,k}^T}/{ \sum_k w_{j,k}^T}$,
and each UE $k$ associated with the BS $j$ is assigned a bandwidth of $ W_{j,k}^{T} = w_{j,k}^T \times W_{\rm tot}$. 

\begin{figure}[!t]
\centering
\subfloat [CDF of SINR] {
	\includegraphics[width=0.45\textwidth]{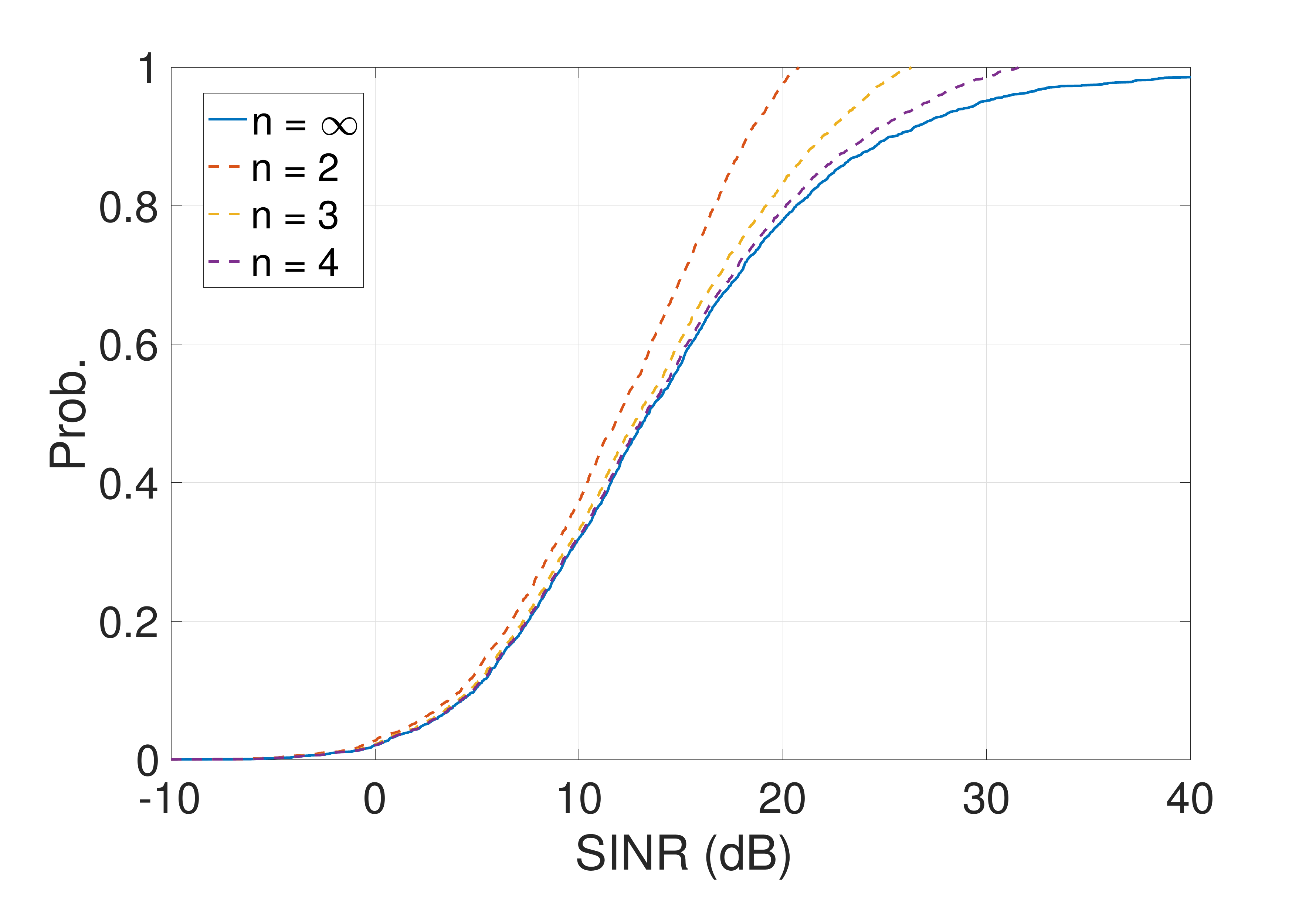} \label{fig:fdmaa}
} \qquad
\subfloat [CDF of Rate] {
	\includegraphics[width=0.45\textwidth]{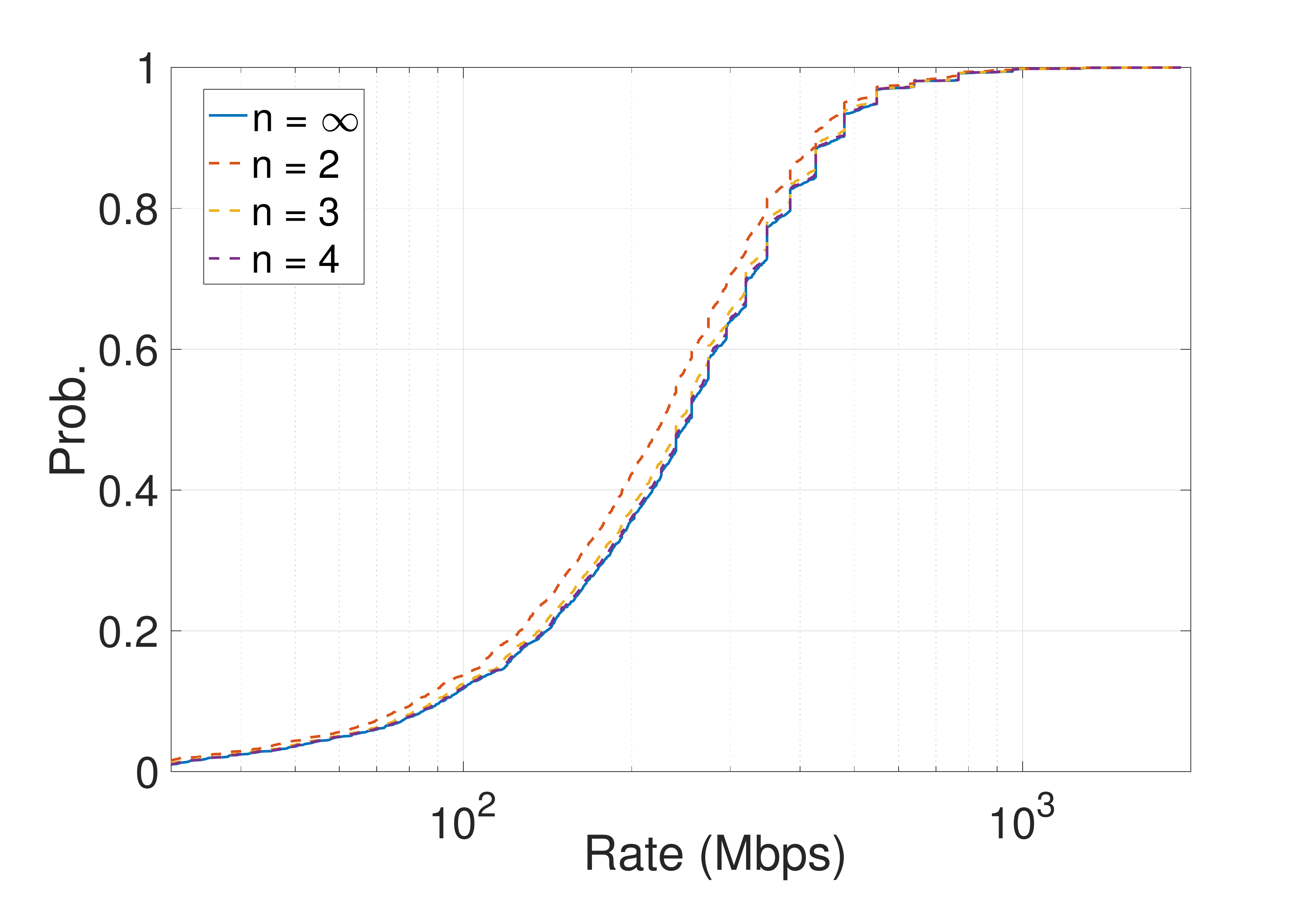} \label{fig:fdmab}
}
\caption{Millimeter wave DL multi user simulations showing the effect of low resolution ADCs on the link quality and achieved rates using OFDMA as the multiple access scheme. For $n\geq3$ the loss due to quantization becomes negligible since very few users operate at a sufficiently high SNR for quantization noise to have noticeable impact.}
\label{fig:fdma}
\end{figure}

In Fig. \ref{fig:fdmaa} we plot the distribution function of the DL SINR for systems with $n = 2, 3$ and $4$ bits of quantization at the Rx. For comparison, we also plot the case when infinite ADC resolution is available ($n=\infty$) at the Rx. We observe that at low SINR, the deviation from the $n=\infty$ curve is minimal if any. On the other hand, at high SINR regimes we observe a ``clipping'' of the maximum achievable SINR. More specifically, the SINR penalty for 2 bit quantization is nearly $10$~dB for the $90$-th percentile UE. For a $3$-bit Rx, the $90$-th percentile UEs have less than $5$~dB of loss in SINR. In the $50$-th percentile on the other hand, this difference goes down to about $2.5$ and $1$~dB for $2$ and $3$-bits of resolutions respectively.   

We next plot the achievable rates under various quantizer resolutions in Fig. \ref{fig:fdmab}. Following the analysis in \cite{AkdenizCapacity:14} and the link-layer model \cite{MogEtAl:07}, we assume a $3~$dB loss from Shannon capacity, and a 20\% overhead. A maximum spectral efficiency of $\rho = 7.4063$ b/s/Hz is assumed based on the 256 QAM modulation scheme proposed in 3GPP NR standards \cite{3GPPTS38214}. The loss due to quantization is not noticeable for $n \geq 3$. This is because very few users in the system will operate at high SINR, thus the clipping of SINR as observed in Fig. \ref{fig:fdmab} has little effect on the average rate. Moreover, as rate is a logarithmic function of the SINR, increasing the SINR beyond a certain point produces diminishing increase in the rate, more so with a limit on the maximum spectral efficiency.
Further, we observe that under full buffer assumption, TDMA and OFDMA will achieve same rates under identical settings. But, OFDMA is only possible through digital BF, and is more efficient for low latency transmission of short mission critical data packets. Analog or hybrid BF based systems have to rely on TDMA where transmission of short packets can be either wasteful in terms of radio resource utilization or incur high latencies \cite{dutta2017frame}.

\paragraph*{SDMA} Following our discussion in Sec. \ref{secMA}, for SDMA systems, each BS assigns the entire bandwidth to $k$ users at any given TTI. The number of users scheduled at each TTI depends on the maximum number of simultaneous beams supported by the system $N_{\rm beam}^{\rm max}$ and the multi-user channel condition. A simple scheduler based on proportional fair selection and sum rate maximization is used to demonstrate our results. 
At the $T$-th TTI, the $j$-th BS will select a group of UEs $\mathcal{K}(T)$ where $|\mathcal{K}(T)| \leq N_{\rm BS}^{\rm beams}$. The first UE $k_1$ is selected into the scheduled group $\mathcal{K}(T)$ such that,
\beq 
w_{j, k_1}^T = \max_{k} w_{j,k}^T,
\eeq 
where $w_{j,k}^T$ is computed using (\ref{eq:PFwt}). Next, the $j$-th BS will admit users to the scheduled group if the achievable sum rate of the BS increases by that admission. The BS will stop admitting users to the scheduled group at a given TTI when either $N_{\rm BS}^{\rm beams}$ UEs are scheduled or given the associated UEs, no UE can be added to the group such that the sum-rate increases. 

\begin{figure} [!t]
\centering
\subfloat [CDF of SINR for $N_{\rm BS}^{\rm beams} = 2$] {
	\includegraphics[width=0.45\textwidth]{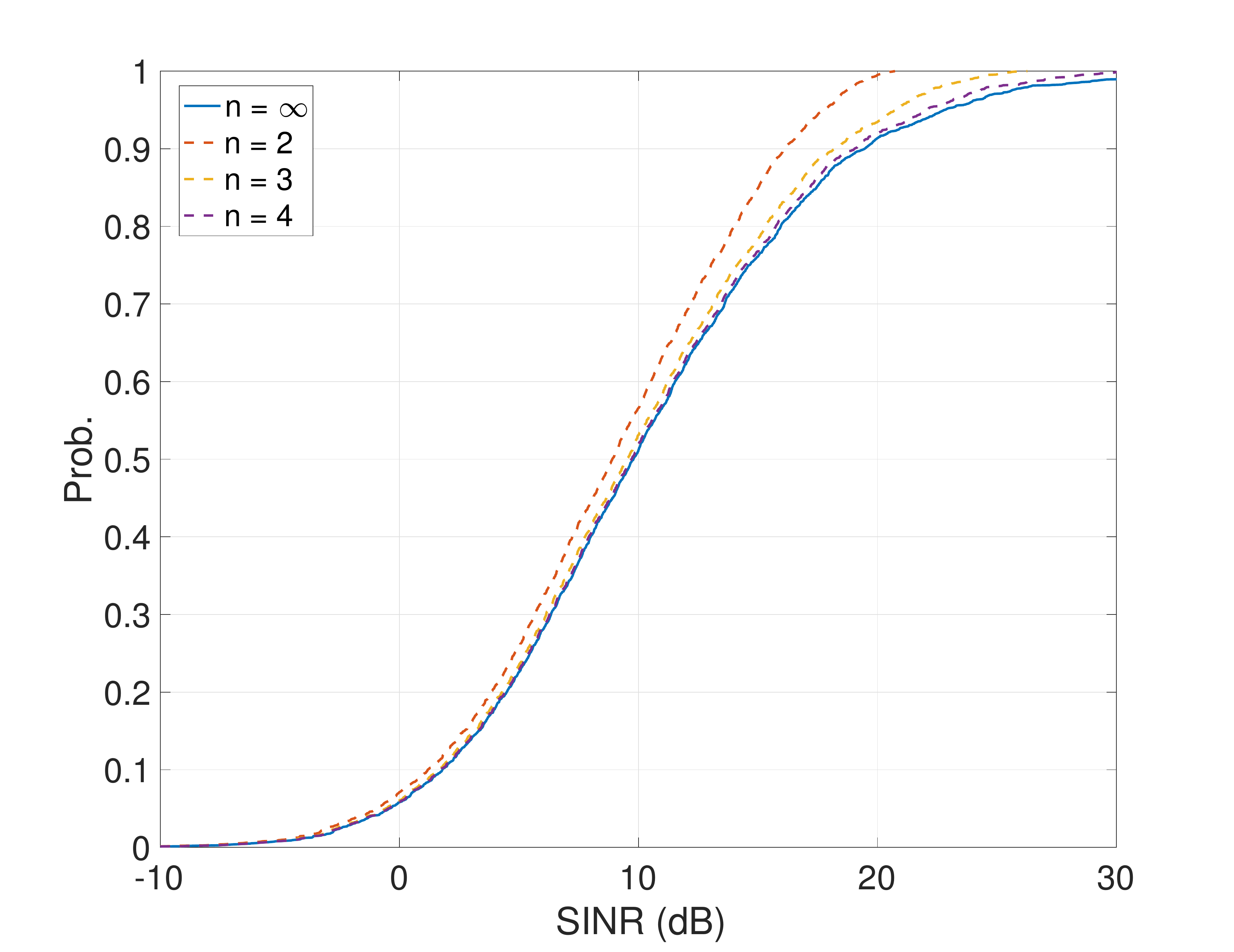} \label{fig:sdma_n2sinr}
} \qquad
\subfloat [CDF of SINR for $N_{\rm BS}^{\rm beams} = 4$] {
	\includegraphics[width=0.45\textwidth]{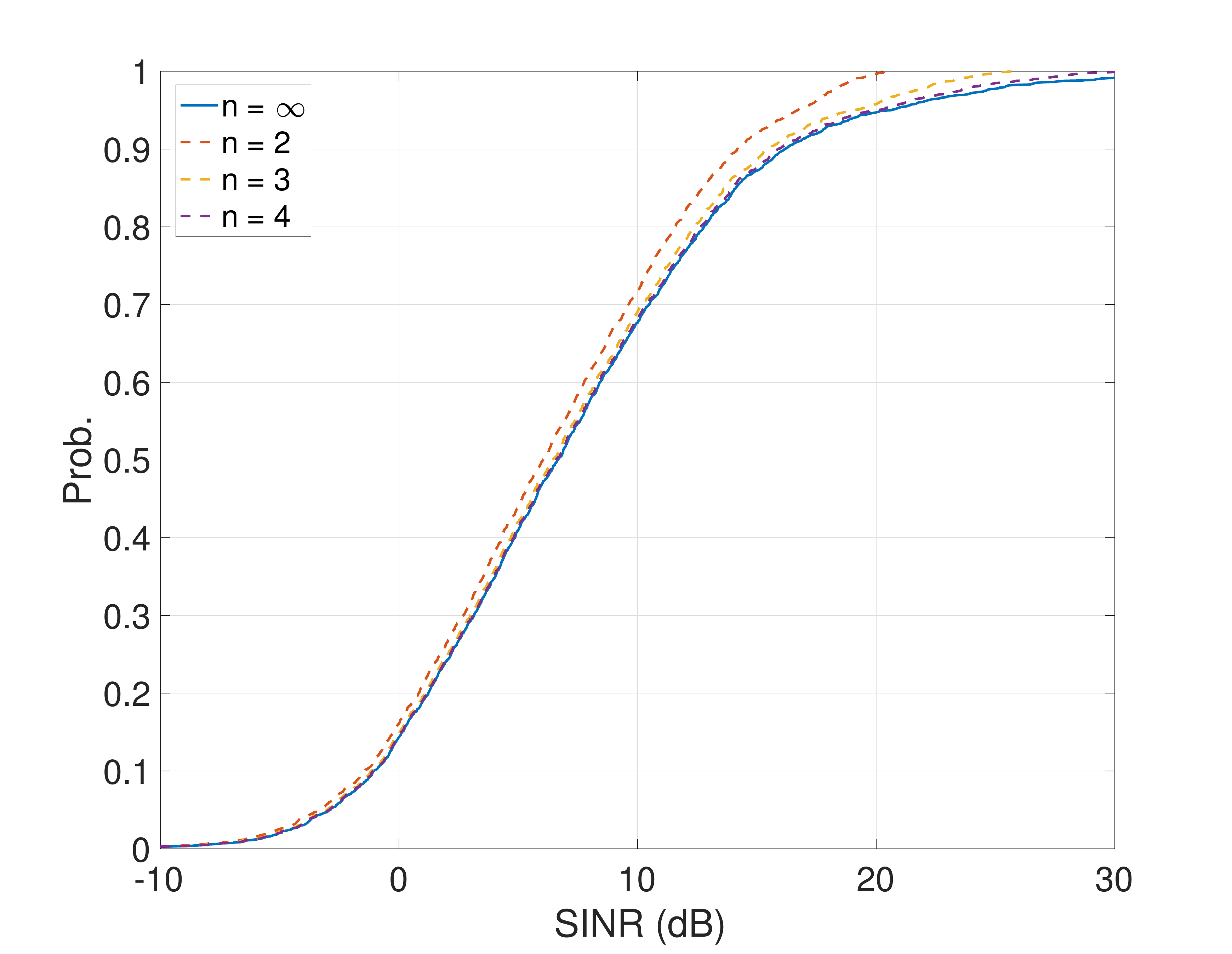} \label{fig:sdma_n4sinr}
}

\subfloat [CDF of rate] {
	\includegraphics[width=0.45\textwidth]{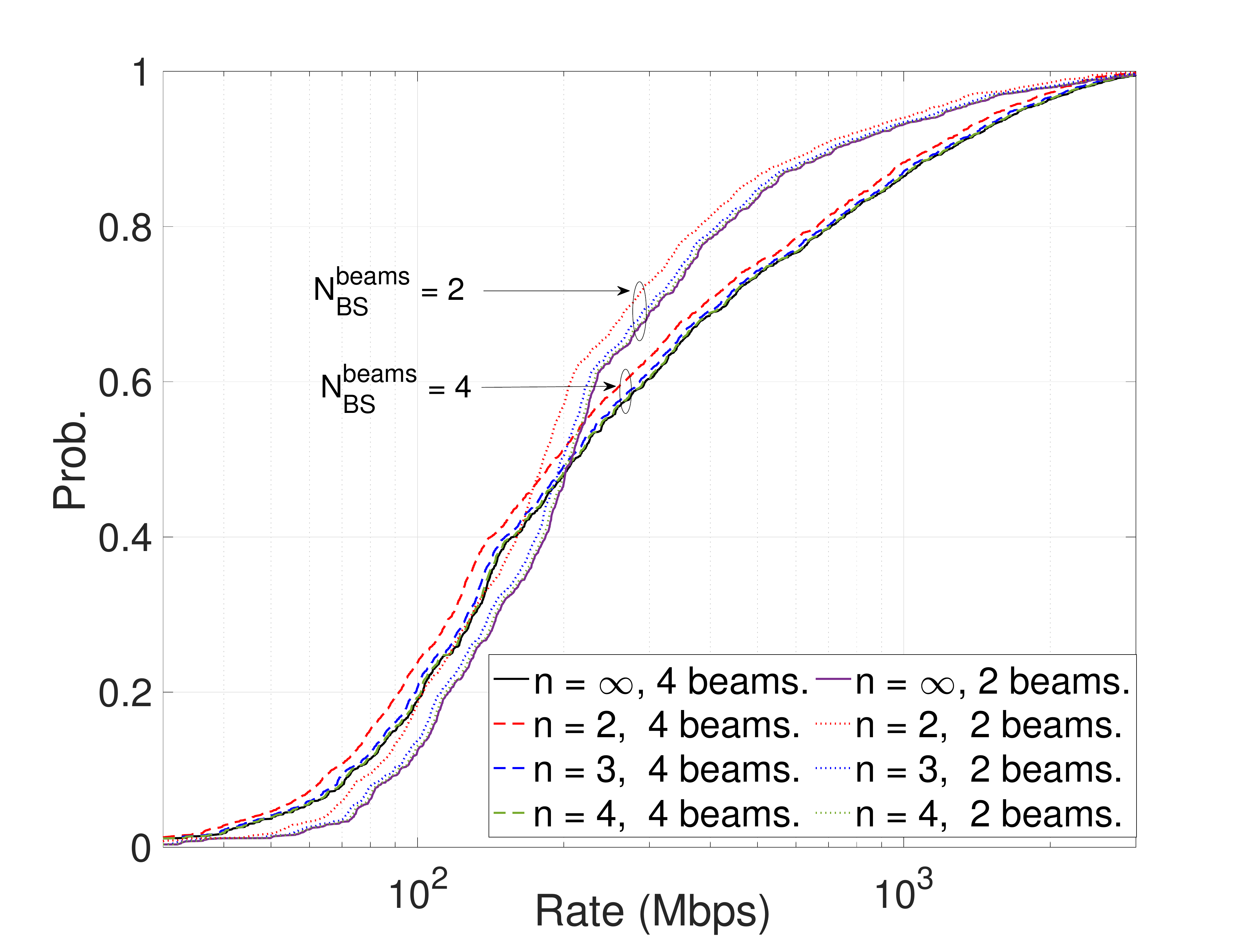} \label{fig:sdma_rate}
} \qquad
\subfloat [Distribution of the number of UEs scheduled by each BS per TTI] {
	\includegraphics[width=0.45\textwidth]{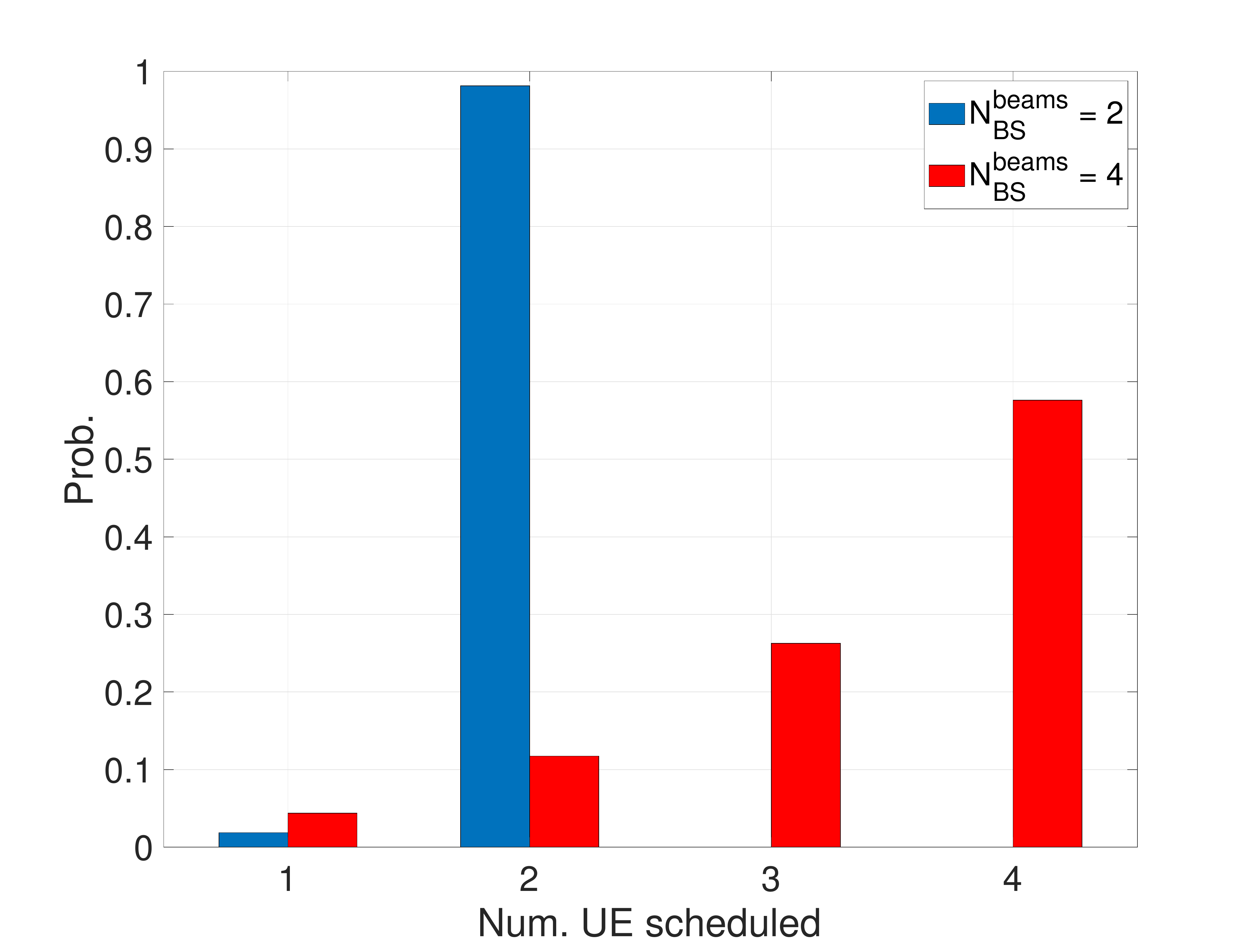} \label{fig:sdma_sched}
}
\caption{Millimeter wave DL multi user simulations showing the effect of low resolution ADCs on the link quality and achieved rates for SDMA with $N_{\rm BS}^{\rm beams} = \{2, 4\}$ spatial streams. }
\label{fig:sdma_cellular}
\end{figure}

We present our results for a SDMA system with $ N_{\rm BS}^{\rm beams} = 2$ and $4$ in Fig. \ref{fig:sdma_cellular}. From Fig. \ref{fig:sdma_cellular} we observe the SINR clipping due to quantization noise. We also notice the effect of ICI on the system by comparing Fig. \ref{fig:sdma_n2sinr} with Fig. \ref{fig:fdmaa}. We observe that with 2-stream SDMA around $5\%$ of the users have a SINR less than $0~$dB; with OFDMA less than 1\% of the UEs have SINR $< 0$~dB. Also, the median SINR in the SDMA case is $10$~dB for $N_{\rm BS}^{\rm beams} = 2$ and $7$~dB for $N_{\rm BS}^{\rm beams} = 4$; a $3$ and $6$ dB additional loss compared to OFDMA with infinite resolution ADCs.
In spite of the presence of ICI, like OFDMA, the effects of low resolution quantization is only noticeable in the high SINR regimes as evident from Figs. \ref{fig:sdma_n2sinr} and \ref{fig:sdma_n4sinr}. We note that due to ICI, the beamformed DL SINR rarely exceeds $30$~dB hence, with $n=4$ or more bits of resolution, there is no noticeable loss in link quality.
Moreover, with multi-stream SDMA, rates greater than $1$ Gbps can be achieved for the top 7\% and 14\% users with $N_{\rm BS}^{\rm beams} = 2$ and $4$ respectively as shown in Fig. \ref{fig:sdma_rate}. This is a considerable improvement over OFDMA where less than the top 1\% of the users achieved rates higher than 1 Gbps. More importantly, we notice that there is very little penalty in the achieved rates due to quantization noise. Especially with $n=3$ or more bits of resolution, the effect of quantization noise at the receiver on the average rate is negligible. 

Interestingly, in Fig. \ref{fig:sdma_sched} we observe that with $N_{\rm BS}^{\rm beams} = 2$, in more than $95\%$ of the scheduling instances, the maximum possible number of beams are used. On the other hand when $N_{\rm BS}^{\rm beams} = 4$, we see that less than $60\%$ of the time are all the beams are used. 
%This points to that fact that novel scheduling algorithms are required to efficiently utilize the spatial degrees of freedom offered by digital beamfoming by reducing ICI. 
Based on our results, we draw the following conclusions. To fully utilize the available spatial degrees of freedom offered by fully digital beamforming for data transmissions, sophisticated scheduling algorithms will be necessary. Yet, even when large number of users are scheduled simultaneously, the effect of low resolution quantization is negligible in the case of SDMA transmissions.
%even when a large number of streams are simultaneously utilized by the mmWave transmitter, the loss in rate due to low resolution quantizers is negligible; the main source of loss is ICI for SDMA systems.

\subsection{Transmitter Characteristics} \label{sec:6c}

\begin{figure}[!t]
\centering
\includegraphics[width=0.65\textwidth]{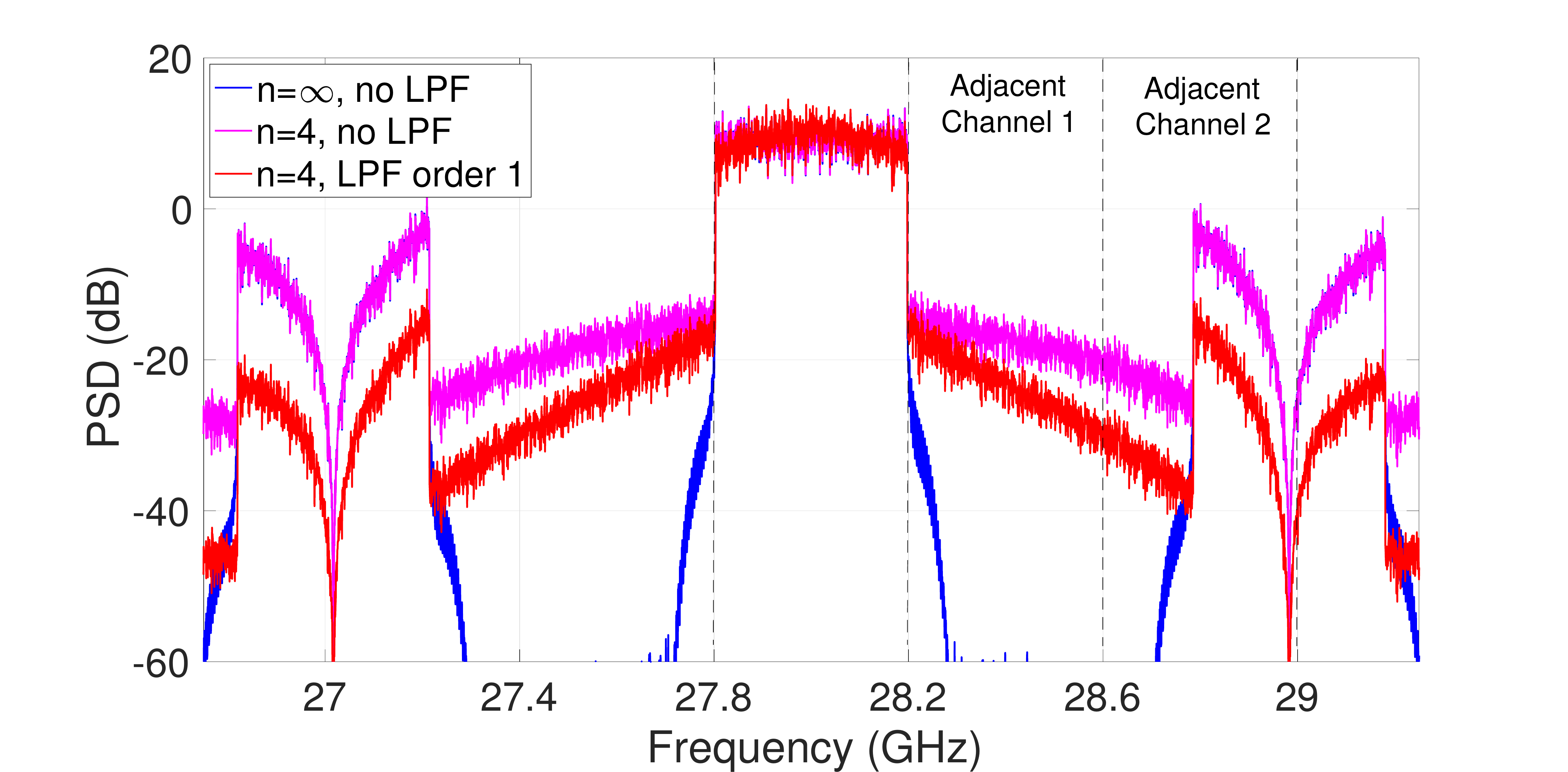} 
\caption{Transmit p.s.d of a 400 MHz OFDM signal using a DAC sampled at $f_s = 983.04$~MHz centered at $f_c =28$ GHz. Observe the effects of quantization noise on the adjacent channels centered at $28.4$ and $28.8$ GHz where a LPF of order 1 has considerable impact on leakage attenuation.}
\label{fig:psd}
\end{figure}

Finally, we look into the effects of a low resolution DAC on the transmitted signal quality. As discussed in Sec. \ref{Sec5}, the quantization noise not only corrupts the transmitted signal but also increases the leakage into the adjacent channels. This is evident from the power spectral density (p.s.d) of the transmitted signal plotted in Fig. \ref{fig:psd}. The signal is transmitted over a $f_{\rm BW}^{\rm ch}=400$~MHz channel around $f_c = 28$~GHz. The adjacent channels are $f_{\rm BW}^{\rm ch}$ wide and located around $f = f_c \pm n f_{\rm BW}^{\rm ch},~ n = 1,2,\ldots$.  The DAC operates at a sampling frequency of $f_s = 2\times f_{\rm chip} = 983.04$~MHz. The ACLR is measured over the maximum occupied bandwidth of $f_{\rm BW}^{\rm meas} = 396$~MHz.

\begin{figure}[!t]
\centering
\subfloat [ACLR on adjacent channel 1. ] {
	\includegraphics[width=0.45\textwidth]{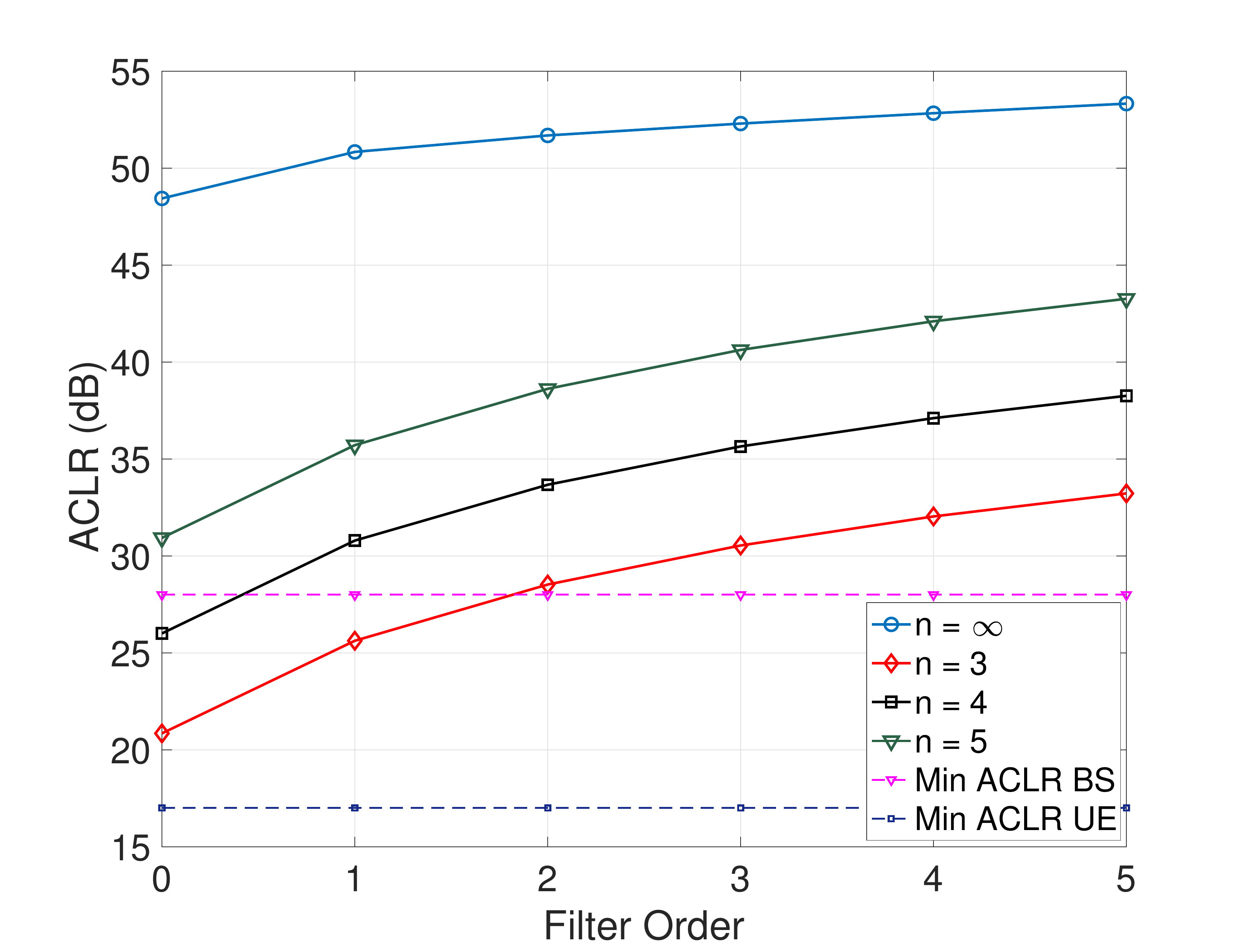} \label{fig:aclr1}
} \qquad
\subfloat [ACLR on adjacent channel 2.] {
	\includegraphics[width=0.45\textwidth]{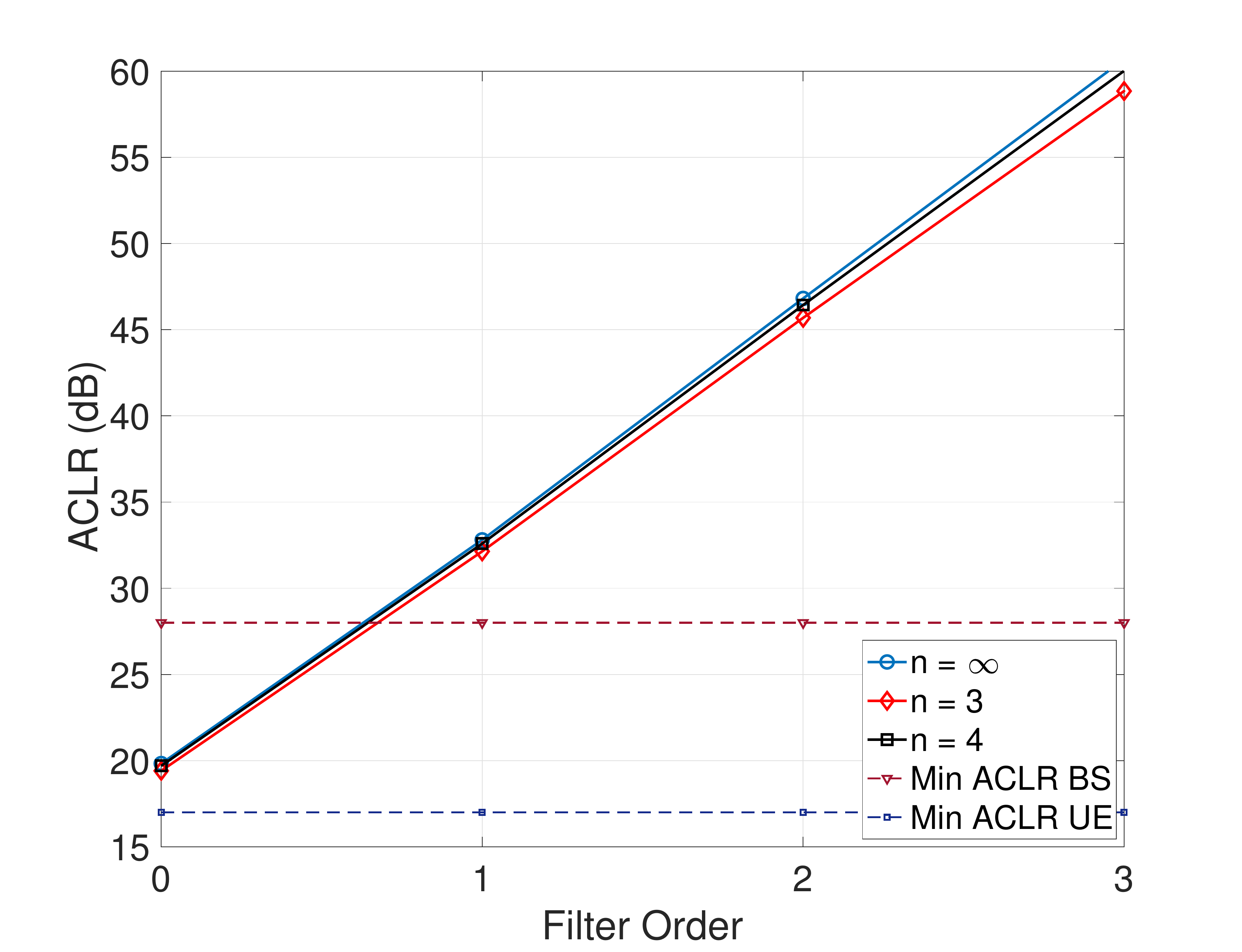} \label{fig:aclr2}
}
\caption{ACLR versus filter order for Butterworth LPF measured over adjacent channels 1 and 2. A filter order of 0 implies the absence of a LPF at the DAC output. Also shown in dashed lines are the ACLR requirements at the BS and the UE specified by 3GPP NR specifications at mmWave frequencies (frequency range 2).}
\label{fig:aclr}
\end{figure}

In Fig. \ref{fig:psd} we can note two important points. Firstly, the quantization noise considerably increases the leakage in the adjacent channels. For instance, the leakage is nearly 40 dB higher for $n=4$ compared to the $n=\infty$. Secondly, the LPF serves two crucial purposes. Not only does it attenuate the out-of-band quantization noise, it also removes the spectral images introduced by the DAC. In fact, with $n=\infty$, adjacent channel 2 in Fig. \ref{fig:psd}, has a very low ACLR due to the presence of the spectral image. In Fig. \ref{fig:aclr} we plot the ACLR on adjacent channel 1 and 2 versus the LPF order when the LPF is modeled as a Butterworth filter. For 5G mmWave systems, the 3GPP standards \cite{3GPPTS38104, 3GPPTS381012} specify the ACLR to be $28$~dB and $17$~dB for BSs and UEs respectively. From Fig. \ref{fig:aclr1} we see that for $n \geq 4$, an order-1 Butterworth filter is sufficient to meet the ACLR requirements at the BS.  Moreover, from Fig. \ref{fig:aclr2}, we note that to achieve acceptable ACLR over adjacent channel 2, i.e., have sufficient image rejection, the BS Tx requires at least an order-1 Butterworth LPF. 

Interestingly, we observe that at the BS, in order to knock out spectral images, a Butterworth LPF of order 1 is necessary regardless of the DAC resolution. Furthur, this LPF is also sufficient to attenuate the out of band quantization noise below the level specified in \cite{3GPPTS38104} when $n \geq 4$ bits of quantization are used. Thus, for a BS Tx, the out of band emissions due to finite quantization can be sufficiently attenuated without an increase in the hardware complexity when compared to the infinite resolution case. Moreover, as shown in Sec. \ref{Sec2:circts}, low order active analog filters consume very little power in the current state of the art. For UEs, the ACLR requirements \cite{3GPPTS381012} are met for $n \geq 3$ without any assumption on filtering for both adjacent channels 1 and 2. Hence, low resolution DACs can be used on UE FEs possibly without any analog LPF. For fully digital systems, this implies a saving in power or chip area for UEs.

\begin{figure}[!t]
\centering
\includegraphics[scale=0.2]{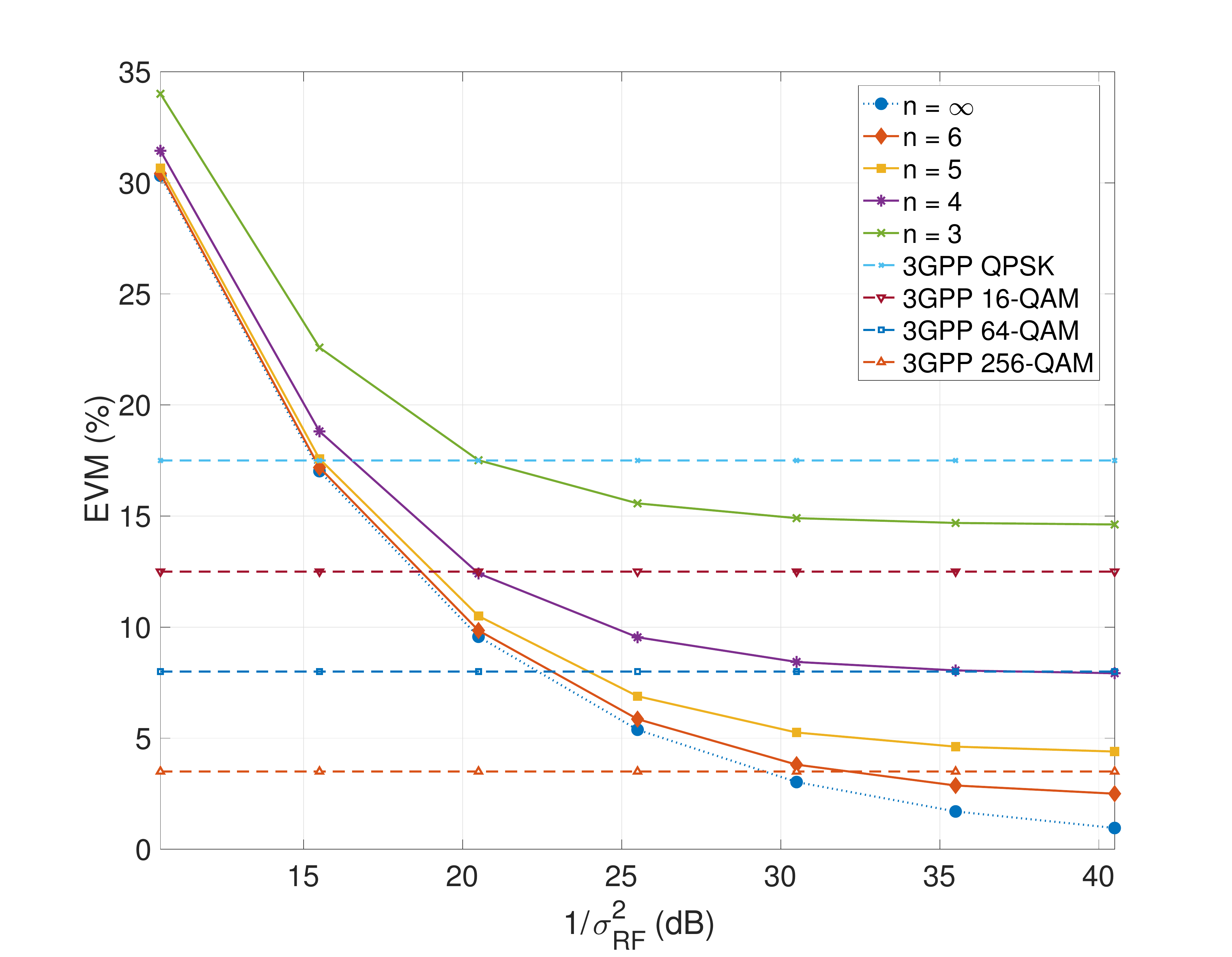} 
\caption{EVM vs. RF impairment comparing the performance of finite resolution DACs with a first order Butterworth LPF at the output. Dashed lines represent the minimum EVM requirements for different modulation orders.}
\label{fig:evm}
\end{figure}
To conclude our study on low resolution DACs, in Fig. \ref{fig:evm} we plot the EVM versus $1/\sigma_{RF}^2$ representing the Tx signal to RF impairments ratio as discussed in Sec. \ref{sec:EvmTh} when the Tx power is normalized to unity. %The DAC has an order-1 Butterworth LPF at its output.
From Fig. \ref{fig:evm} we note that $4$-bits of resolution are sufficient to support modulation orders up to 64-QAM which, as specified in \cite{3GPPTS38214}, is the highest modulation order that UEs need to support. Thus, 4 bits of resolution is sufficient for DACs used in UEs. For the BS transmitter, using $n \leq 5$ implies that $256$-QAM cannot be supported due to quantization noise. This does not violate specification, as $256$-QAM is an optional feature in current NR standards, but limits the maximum achievable spectral efficiency on the DL. Thus, to fully exploit the large bandwidth at mmWave, the BS transmitter will need DACs with at least $6$-bits of resolution. This is a feasible design choice for BSs as they operate with higher power budgets than UEs.

\section{Conclusion}\label{SecConc}
Fully digital beamforming at mmWave requires the use of low resolution converters to keep the power consumption of the front ends reasonable. The gain in spatial multiplexing offered by the fully digital architecture is, thus, achieved at a cost of signal degradation due to coarse quantization. In this paper we have determined how many bits of resolution is required for efficient communications over wide band mmWave channels. We show that at the Rx, the loss due $3-4$ bits of ADC resolution is negligible for practical cellular deployment scenarios. More interestingly, we show that mmWave receivers with $3-4$ bits of ADCs precision can achieve multi-Gbps rates when SDMA scheduling is used for medium access.
For transmitters, low resolution DACs, with $4$ or more bits of precision, meet the 3GPP transmission regulations on ACL without any additional hardware costs both at the BS and and the UE.
Further, we show that the EVM required for the transmission of $64$-QAM is met by $4$-bit DACs while $6$-bits of DAC resolution is required to support $256$-QAM. This implies that $4$-bits of resolution is sufficient at the UE Tx while $6$-bits may be required for BS DACs when the $256$-QAM is to be supported. Thus, low resolution fully digital beamforming can be used both at the receiver and the transmitter of wide band mmWave cellular equipments.

%this needs some work!
A major concern for fully digital beamformers at mmWave will be the cost and power draw by the baseband processor. This requires the design and analysis of digital beamformers at mmWave, similar to \cite{yang2018}, with low resolution converters. Moreover, the flexibility offered by digital beamforming cannot be exploited without efficient communication protocols. Current mmWave cellular systems are designed on the assumption of analog or hybrid beamforming. To enable ultra low latency communications using fully digital BF, the design of control and data channels need to be revisited at mmWave frequencies. In fact, future research and standardizations efforts need to consider the practicality and potentials of fully digital beamforming.

\bibliographystyle{IEEEtran}
\bibliography{bibl}

\end{document}